\documentclass[twocolumn,nofootinbib]{revtex4}
\usepackage{graphicx}
\usepackage{latexsym}
\def\be{\begin{equation}}
\def\ee{\end{equation}}
\def\bea{\begin{eqnarray}}
\def\eea{\end{eqnarray}}

\begin{document}

\title{Scale dependence of the primordial spectrum from combining the three-year WMAP,
Galaxy Clustering, Supernovae, and  Lyman-alpha forests }
\author{Bo Feng${}^{a}$}

\author{Jun-Qing Xia${}^b$}

\author{Jun'ichi Yokoyama${}^{a}$}

\affiliation{ ${}^a$ Research Center for the Early Universe
(RESCEU), Graduate School of Science, The University of Tokyo, Tokyo
113-0033, Japan}

\affiliation{${}^b$Institute of High Energy Physics, Chinese Academy
of Science, P.O. Box 918-4, Beijing 100049, P. R. China}

\date{\today.}
\begin{abstract}

We probe the scale dependence of the primordial spectrum in the
light of the three-year WMAP (WMAP3) alone and WMAP3 in combination
with the other cosmological observations such as  galaxy
clustering and Type Ia Supernova (SNIa). We pay particular attention
to the combination with the Lyman $\alpha$ (Ly$\alpha$) forest. Different
from the first-year WMAP (WMAP1), WMAP3's preference on the running
of the scalar spectral index on the large scales is now fairly
independent of the low CMB multipoles $\ell$. A combination with the
galaxy power spectrum from the Sloan Digital Sky Survey (SDSS)
prefers a negative running to larger than 2$\sigma$, regardless the
presence of low $\ell$ CMB ($2\le \ell \le 23$) or not. On the other hand
if we focus on the Power Law $\Lambda$CDM cosmology with only six
parameters (matter density $\Omega_m h^2$, baryon density $\Omega_b
h^2$, Hubble Constant $H_0$, optical depth  $\tau$, the spectral index,
 $n_s$,
and the amplitude, $A_s$, of the scalar perturbation spectrum)
 when we drop the low $\ell$ CMB contributions WMAP3 is
consistent with the Harrison-Zel'dovich-Peebles scale-invariant
spectrum ($n_s=1$ and no tensor contributions) at $\sim 1\sigma$.
 When assuming a
simple power law primordial spectral index or a constant running, in
case one drops the low $\ell$ contributions ($2\le \ell \le 23$) WMAP3 is
consistent with the other observations better, such as the
inferred value of $\sigma_8$.
We also find, using a spectral shape with a minimal extension
of the running spectral index model,
LUQAS$+$ CROFT Ly$\alpha$  and SDSS
Ly$\alpha$ exhibit somewhat different
preference on the spectral shape.

\end{abstract}

\maketitle

\hskip 1.6cm PACS number(s): 98.80.Es, 98.80.Cq \vskip 0.4cm

\section{Introduction}

It is commonly believed that the early Universe has experienced a
stage of accelerated expansion which is known as
inflation \cite{Sato:1980yn,Guth:1980zm,Linde:2005ht}. During
inflation one of the tiny patches of the very early Universe
was superluminally stretched to become our observable Universe
today. The theory of inflation can naturally explain why the
universe is flat, homogeneous and isotropic. Inflation is driven by
the potential energy of a scalar field called inflaton and its
quantum fluctuations turn out to be the primordial density
fluctuations which seed the observed large-scale structures (LSS)
and anisotropy of the Cosmic Microwave Background (CMB) Radiation.
Inflation typically ends with a period of reheating when the
inflaton starts rapid oscillations and it decays into standard
particles. Then the universe starts the standard hot Big Bang
evolution.

Over the past decade, the theory of inflation has successfully
passed several nontrivial tests. In particular, the first year
Wilkinson Microwave Anisotropy Probe(WMAP)
 \cite{Bennett:2003bz,Kogut:2003et,Peiris:2003ff,Spergel:2003cb,Hinshaw:2003ex,Verde:2003ey}
has detected a large-angle anti-correlation in the
temperature-polarization cross-power spectrum, which is the
signature of adiabatic superhorizon fluctuations at the time of
decoupling \cite{Peiris:2003ff}.

The recently released three year
WMAP (WMAP3)
 \cite{Spergel:2006hy,Page:2006hz,Hinshaw:2006,Jarosik:2006,WMAP3IE}
has marked another milestone on the precision cosmology of CMB
Radiation. The simplest six-parameter power-law $\Lambda$CDM
cosmology, namely with matter density $\Omega_m h^2$, baryon density
$\Omega_b h^2$, Hubble Constant $H_0$, optical depth  $\tau$, the
spectral index and amplitude of the scalar perturbation spectrum,
$n_s$ and $A_s$, turn out to be in good agreement with WMAP3
together with the small-scale CMB observations such as
BOOMERANG  \cite{Montroy:2005yx}, the Arcminute Cosmology Bolometer
Array Receiver (ACBAR  \cite{Kuo:2002ua}), the Cosmic Background
Imager (CBI  \cite{Readhead:2004gy}) and the Very Small
Array(VSA  \cite{Dickinson:2004yr}), LSS as measured by Two degree
Field (2dF)  \cite{Cole:2005sx} and the Sloan Digital Sky Survey
(SDSS)  \cite{Tegmark:2003uf} and with the Type Ia Supernova (SNIa)
as measured by the Riess ``gold'' sample  \cite{Riess:2004nr} and the
first year Supernova Legacy Survey (SNLS)  \cite{snls}. This
agreement between the above ``canonical'' cosmological model and
observations can be used to test a number of possible new physics,
such as the equation of state of dark energy, neutrino masses,
cosmic CPT violation, etc.

Although the temperature-temperature correlation power (TT) of
WMAP3 is now cosmic-variance limited up to $\ell \sim 400$  and the
third peak is now detected, the features discovered by the first
year WMAP \cite{Peiris:2003ff,Spergel:2003cb} are still present: the
low TT quadrupole and localized oscillating features on TT for $\ell
\sim 20-40$ \cite{Hinshaw:2006}. Although the signatures of glitches
on the first peak discovered by the first year WMAP have now become
weaker, they do exceed the limit of cosmic
variance \cite{Hinshaw:2006}.  In fact, even before the release of the
first year WMAP Ref.  \cite{Wang:2002hf} claimed oscillating
primordial spectrum could lead to oscillations around the first peak
of CMB TT power. These data  have hence revealed many
interesting detailed features which cannot be explained in the
simplest version of the inflation model, and we are making much
effort to construct a realistic particle physics model of inflation.

As for the shape of the primordial power spectrum, it is also
noteworthy that a significant deviation has been observed by WMAP3
from the simplest Harrison-Zel'dovich-Peebles scale-invariant
spectrum ($n_s=1$ and no tensor contributions), and that this
feature is more eminent with the combination of all the currently
available CMB, LSS and SNIa (dubbed the case of ``All'' in
 \cite{WMAP3IE}). On the other hand, a nontrivial negative running
of the scalar spectral index $\alpha_s$, whose existence was studied
even before WMAP epoch \cite{Feng:2003nt}(for relevant study see
also  \cite{Lewis:2002ah}), was favored by the first-year WMAP
papers  \cite{Bennett:2003bz,Peiris:2003ff,Spergel:2003cb}. But its
preference was somehow diminished as corrections to the likelihood
functions were made
 \cite{Cline:2003ve,Efstathiou:2003tv,Slosar:2004fr}. Dramatically,
the new WMAP3 data prefers again a negative running in the ``All''
combination  \cite{WMAP3IE}. If  confirmed, a nonvanishing  running
of $\alpha_s$ would not only constrain inflationary cosmology
significantly
\cite{Feng:2003mk,Kawasaki:2003zv,Huang:2003zp,Yamaguchi:2003fp,Yamaguchi:2004tn,Chen:2004nx,Easther:2006tv,Yokoyama:1998rw,Yokoyama:1998pt},
but also affect the cosmological constraint on the neutrino mass
\cite{Feng:2006zj,toappear}.

In the present paper, using the Markov Chain Monte Carlo (MCMC)
method, we aim to probe the scale dependence of the primordial
spectrum in the light of WMAP3 alone and that in combination with
the other cosmological observations  from galaxy clustering and
SNIa. We pay particular attention to the combination with the Lyman
$\alpha$ (Ly$\alpha$) forest.

We will show that, different from the
first-year WMAP (WMAP1), WMAP3's preference on the running of the
scalar spectral index on the large scales is now fairly independent
of the low CMB multipoles $\ell$. We  also find that a combination
of WMAP3 with the galaxy power spectrum from SDSS prefers a negative
running with more than 2$\sigma$, regardless of
the presence of low $\ell$
CMB ($2\le \ell \le 23$).  On the other hand, if we focus on the
Power Law $\Lambda$CDM cosmology with only six parameters, we
find that when one drops the low $\ell$ CMB contributions WMAP3 would
be consistent with the Harrison-Zel'dovich-Peebles scale-invariant
spectrum at $\sim 1\sigma$. Assuming a
simple power-law primordial spectral index or a constant running, in
cases one drops the low $\ell$  contributions
 ($2\le \ell \le 23$) WMAP3
is consistent with the other observations better, such as
 the value of $\sigma_8$. 

On the other hand, using a spectral shape with a minimal extension
of the running spectral index model, we find
LUQAS$+$ CROFT Ly$\alpha$ \cite{Viel:2004bf} and SDSS
Ly$\alpha$ \cite{McDonald:2004xn} exhibit somewhat different
preference on the spectral shape, although it is difficult to
draw any definite conclusion because the result depends on how
we model the spectral shape.

The rest of the paper is structured as follows. In Section
II we describe the method and the data. In Section III we analyze
 the effects of low CMB multipoles on
the determination of cosmological parameters using WMAP3
 \cite{Spergel:2006hy,Page:2006hz,Hinshaw:2006,Jarosik:2006,WMAP3IE},
SNIa  \cite{Riess:2004nr,snls}, 2dF \cite{Cole:2005sx} and Sloan
Digital Sky Survey 3-D power spectrum (SDSS P(k))
 \cite{Tegmark:2003uf}, LUQAS+CROFT sample of Lyman
$\alpha$ \cite{Viel:2004bf} and SDSS Lyman $\alpha$
sample \cite{McDonald:2004xn}  by global fittings using the MCMC
technique.
In Section IV scale dependence of fluctuations is analyzed with priors
on the Hubble parameter, $\Omega_bh^2$, and the cosmic age.

Discussions and conclusions are presented in the last
section.

\section{Method and data}

To break the possible degeneracies among the variations of the
current cosmological parameters, we make a global fit to the
cosmological observations with the publicly available Markov Chain
Monte Carlo package \texttt{CosmoMC} \cite{Lewis:2002ah,IEMCMC}. We
assume purely adiabatic initial conditions, and impose the flatness
condition motivated by inflation. Our most general parameter space
is
\begin{equation}\label{para}
    \textbf{p}\equiv(\omega_{b}, \omega_{c}, \Theta_S, \tau, n_{s},
     n'_{s}, \alpha_{s}, A_s)~,
\end{equation}
where $\omega_{b}=\Omega_{b}h^{2}$ and $\omega_{c}=\Omega_{c}h^{2}$
are the physical baryon and cold dark matter densities relative to
the critical density, $\Theta_S$ characterizes the ratio of the
sound horizon to the angular diameter distance at decoupling, $\tau$
is the optical depth and $A_{s}$ is defined as the amplitude of the
primordial power spectrum at $k=0.05$ Mpc$^{-1}$. The pivot scales
for $n_{s}$  and $\alpha_{s}$ are also chosen at $k=0.05$
Mpc$^{-1}$. Here $h$ is the Hubble constant in unit of 100 km
s$^{-1}$ Mpc$^{-1}$. The bias factors of 2dF, SDSS and LUQAS+CROFT
sample of Lyman $\alpha$ (Ly$\alpha$) as analyzed in
Ref.\  \cite{Viel:2004bf} have been used as nuisance parameters.

For some of our simulations we have included the following priors:
For $h$, we make use of the Hubble Space Telescope (HST)
measurement by multiplying the likelihood function with a Gaussian
centered around $h=0.72$ and with a standard deviation $\sigma =
0.08$ \cite{Freedman:2000cf}. We impose a weak Guassian prior on the
baryon density from Big Bang nucleosynthesis (BBN)
 \cite{Burles:2000zk}: $\Omega_b h^2 = 0.022 \pm 0.002$ (1 $\sigma$).
Simultaneously we will also use a cosmic age tophat prior as 10
Gyr$<t_{0}<$20 Gyr.

In our calculations we have taken the total likelihood to be the
products of the separate likelihoods (${\bf \cal{L}}_i$) of CMB,
LSS, SNIa and Ly$\alpha$. In other words defining
$\chi_{L,i}^2 = -2 \log {\bf
\cal{L}}_i$, we get
\be
\chi^2_{L,total} = \chi^2_{L,CMB}
+ \chi^2_{L,LSS}
+ \chi^2_{L,SNIa} + \chi^2_{L,Ly\alpha}~ .
\ee
If the likelihood function is Gaussian, $\chi^2_{L}$ coincides
with the usual definition of $\chi^2$ up to an additive constant
corresponding to the logarithm of the normalization factor of
${\cal L}$.
In the computation of CMB we
have included the three-year WMAP (WMAP3) data with the routine
for computing the likelihood supplied by the WMAP team
 \cite{WMAP3IE}. For the purpose of comparisons in some cases we also
include the first-year temperature and polarization data
 \cite{Bennett:2003bz,Hinshaw:2003ex} with the routine for computing
the likelihood supplied by the WMAP team  \cite{Verde:2003ey}. To be
conservative but more robust, in the fittings to the 3D power
spectrum of galaxies from the SDSS \cite{Tegmark:2003uf} we have used
the first 14 bins only, which are supposed to be well within the
linear regime  \cite{sdssfit}. In the calculation of the likelihood
from SNIa we have marginalized over the nuisance
parameter  \cite{DiPietro:2002cz}. The supernova data we use are the
``gold'' set of 157 SNIa published by Riess $et$ $al$ in
Ref.\  \cite{Riess:2004nr} and the 71 high redshift type Ia supernova
discovered during the first year of the 5-year Supernova Legacy
Survey (SNLS)  \cite{snls}, respectively. In the fittings to SNLS we
have used the additional 44 nearby SNIa, as also adopted by the SNLS
group  \cite{snls}.  In order to be conservative but more robust, we did
not try to combine SNLS together with the Riess sample for
cosmological parameter constraints\footnote{We thank Chris Lidman
for enlightening discussions on the possible systematics led by
combining Riess sample \cite{Riess:2004nr} directly with the SNLS
sample \cite{snls} in the simple way, and Xiao-Feng Wang for
stimulating discussions on the SNIa correlations as probed by the
BATM method in Ref. \cite{Prieto:2006ex}, MLCS2k2 in Ref.
 \cite{Jha:2002af}, SALT in Ref.  \cite{Guy:2005me} and the color
parameter $\Delta C12$ as a new luminosity indicator in
Refs. \cite{Wang:2005qn,Wang:2006cq}. We also thank Xiao-Feng Wang
for showing us his preliminary results in combining SNLS with the
Riess sample in a more professional way \cite{Wang:06ap}.}.

The Lyman-$\alpha$ forest corresponds to the Lyman-$\alpha$
absorption of photons travelling from distant quasars ($z\sim 2-4$)
by the neutral hydrogen in the intergalactic medium (IGM). It is
theoretically possible to infer the matter power spectrum from the
Lyman-$\alpha$
forest \cite{Lesgourgues:2006nd,Viel:2004bf,McDonald:2004xn,Viel:2006yh,Seljak:2006bg}.
Previously Refs. \cite{Viel:2006yh,Seljak:2006bg} have made some
studies on the shape of the primordial spectrum and $\sigma_8$ using
Ly$\alpha$ data after the release of WMAP3. Here we use the LUQAS+CROFT
sample of Lyman $\alpha$ as analyzed in Ref. \cite{Viel:2004bf} (
dubbed V-Ly$\alpha$) and the SDSS Lyman $\alpha$ sample from
Ref. \cite{McDonald:2004xn} (dubbed S-Ly$\alpha$) for our study. We do not
try to combine the two sets of Ly$\alpha$ simultaneously given the possibly
unknown systematics of Ly$\alpha$ \cite{Lesgourgues:2006nd}.

The likelihood of V-Ly$\alpha$ \cite{Viel:2004bf} has been included in the
publicly available package
\texttt{CosmoMC} \cite{Lewis:2002ah,IEMCMC}, which can be used
directly to probe some scale dependence of the primordial spectrum.
On the other hand for S-Ly$\alpha$ analyzed in Ref. \cite{McDonald:2004xn}
we need to use the online C++ code \cite{IESlya1}. The package of
\texttt{CosmoMC} being written in Fortran 90, it is not so
straightforward to apply the code in Ref. \cite{IESlya1} in
\texttt{CosmoMC}. Recently Slosar \cite{IESlya2} made a patch of SDSS
Ly$\alpha$ to \texttt{CosmoMC}. We have corrected the patch \cite{IESlya2}
after some nontrivial crosschecks and made our analysis for S-Ly$\alpha$
basing on the corrected patch.

Regarding the first-year WMAP data the E mode polarization not being
directly measured, there have been simultaneously low TT multipoles
and high TE
multipoles \cite{Bennett:2003bz,Kogut:2003et,Hinshaw:2003ex}. To be
conservative Ref. \cite{Contaldi:2003zv} used only TT data to probe
the possible new physics during inflation and Ref. \cite{Dore:2003wp}
showed that if assuming without significant contaminations, in the
general framework the observed TE amplitude on the largest scales
would be in very large discrepancy with the observed TT for the
concordance $\Lambda$CDM model. Moreover Bridle et al.\
 \cite{Bridle:2003sa} found that the claimed preference of a negative
$\alpha_s$ was merely due to the lowest WMAP1 multipoles. In order
to probe the sensitivity of the running to the lower multipoles we
analyze the running properties using the CMB data with and without
the contributions of lower multipoles which suffer from large cosmic
variance \cite{Bond:1987ub}.

For WMAP3 the large scale CMB data not only include the temperature
power spectrum TT, but also the measurements of the polarization
spectrum EE, BB and the temperature polarization spectrum TE. Given
the distinctive feature of low $\ell$
CMB \cite{Efstathiou:2003tv,Slosar:2004fr}, for WMAP3 the pixel-based
method has been applied to $C_{\ell}^{TT}$ for $2\le \ell \le 12$ and to
polarization for $2\le \ell \le 23$ in
Refs.\  \cite{Spergel:2006hy,Page:2006hz,Hinshaw:2006,WMAP3IE}.
These large-scale components suffer from larger cosmic variance,
possibly uncleaned foreground contaminations and their statistical
nature is somewhat different from Gaussian. For these reasons we
have tentatively prepared two different data-sets of WMAP3, one
being the ``normal'' one with the full range TT, TE and other
polarization data, the other being the truncated one where all of
the components with $\ell\le 23$ have been dropped. We analyze these
two data-sets separately in various models and compare with each
other. Also for a clear comparison and for the fact that WMAP1
should exhibit similar properties on low $\ell$ to WMAP3, we also
truncate the TT and TE likelihood of WMAP1 at $\ell=24$. Thus for the
truncated data-set essentially for both WMAP1 and WMAP3 we will use
only the TT and TE CMB spectra on the truncated smaller scales. The
truncated data-set, although with less amount of information, may be
more conservative but more robust compared with the normal one.



While the simplest single field inflation generically predicts
nearly scale-invariant primordial spectrum on the CMB relevant large
scales, the inflaton would run faster in its later epoch and for the
scales probed by WMAP$+$ Ly$\alpha$, an appreciable
 deviation from scale invariance is
possibly present even in the framework of simple single-field
inflation models. On the other hand, even for the simplest inflaton
potentials known for a long time, the resulting primordial spectrum
can be certainly more complicated than usually
expected
\cite{Hodges:1989dw,Chen:2004nx,Yokoyama:1998rw,Yokoyama:1998pt}.

Motivated by these we mainly focus on our fittings to the Power Law
$\Lambda$CDM (PLCDM) Model with a constant primordial spectral index
$n_s$ and Running Spectral Index $\Lambda$CDM (RLCDM) Model with a
constant running $\alpha$. In order to see if models with more
drastic scale-dependence are observationally more plausible, we also
consider a Running Spectral Index $\Lambda$CDM Model with a
step-like $n_s$ divided at $k=0.1$ Mpc$^{-1}$ (SRLCDM) as an
example. In this model, we assume there is a constant primordial
spectral index $n'_s$ with no running  for $k>0.1$ Mpc$^{-1}$. While
the amplitudes of $n_s$ and $n'_s$ are not continuous, we naturally
assure the continuity of the primordial scalar spectrum $P_s(k)$ at
0.1 Mpc$^{-1}$.

The parameter $\sigma_8$, being defined as the amplitude of density
fluctuations at 8 $h^{-1}$Mpc, cannot be determined by
 WMAP3 alone, which
can probe the the shape of the primordial spectrum only up to $k \sim
0.1$ Mpc$^{-1}$ \cite{Spergel:2006hy},
without extrapolating the information of the primordial spectrum to
much smaller scales, except for an exotic case that the Hubble
constant is anomalously small,
which is certainly impossible at the first glance in the
concordance cosmology. On the other hand if we assume NO prior on
the shape of the primordial spectrum and reconstruct the shape of
the primordial spectrum in a model independent way like Cosmic
Inversion \cite{Matsumiya:2001xj,Matsumiya:2002tx,Kogo:2003yb,Kogo:2004vt,Kogo:2005qi},
even the extremely exotic values of $h$ still remain possible, at
least in cases we use only the CMB TT observations. In the present
work we do not pursue the shape of the primordial spectrum in a
model independent way and will study the cosmological implications
on the shape of the primordial spectrum and the amplitude of
$\sigma_8$ with the forementioned method.

It is believed that the secondary effects on CMB like the
Sunyaev-Zel'dovich (SZ) effects and CMB lensing effects are
typically very small and their effects
 are comparable with each other.
One would take the risk of getting biased
if considering only one of them \cite{Lewis:2006ma}. Neglecting both
of them is a plausible way in the determination of the cosmological
parameters for the interest of the current paper \cite{Lewis:2006ma}.
Hence in the present study we have neglected both of the secondary
effects.

For each regular calculation of MCMC, we run 6 independent chains
comprising of 150,000-300,000 chain elements and spend thousands of
CPU hours to calculate on a supercomputer. The average acceptance
rate is about 40\%. To get the converged results, we test the
convergence of the chains by Gelman and Rubin \cite{GR92} criteria
and typically get $R-1$ to be less than 0.05, which is more
conservative than the recommended value $R-1<$0.1\footnote{For the
case with WMAP3 only fitting to the Running Spectral Index
$\Lambda$CDM Model with step-like $n_s$ truncated at $k= 0.1$
Mpc$^{-1}$, which will be shown in the later part of this paper, we
have $R-1 \sim 0.08$.}.

\section{Analysis of current observations with and without  low
multipole CMB contributions}

\subsection{WMAP data alone}

\begin{figure}[htbp]
\begin{center}
\includegraphics[scale=0.8]{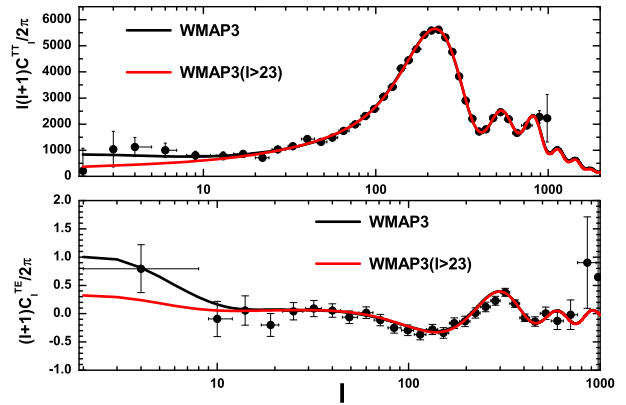}
\caption{ CMB power spectra for the best fit Running Spectral Index
$\Lambda$CDM Model to normal WMAP3 where full range of the data has
been used and to the case where the $\ell \le 23$ data of WMAP3 are
not used. The dots with error bars are binned WMAP3 TT and TE data.
HST, BBN and age priors have been adopted. \label{fig:il}}
\end{center}
\end{figure}

We start with an illustrative figure. In Fig.~ \ref{fig:il} we
delineate the CMB power spectra for the best fit RLCDM Model to
normal WMAP3 where the full range of the WMAP3 observational data
has been useed and to the case where the $\ell \le 23$ data of WMAP3
are not used. The dots with error bars are binned WMAP3 TT and TE
data \cite{WMAP3IE}. HST, BBN and age priors have been adopted. We
have 
$\omega_b=0.0205$, $\omega_c=0.116$, $\Theta_S=1.038$, $\tau=0.092$,
$\log[10^{10} A_s]$ (0.05Mpc$^{-1}$)$=3.039$,
$n_s$(0.05Mpc$^{-1}$)$=0.856$ and
$\alpha_s$(0.05Mpc$^{-1}$)$=-0.0634$ for the best fit case with
normal WMAP3. Correspondingly for the case without small $\ell$
contributions the best fit values are $\omega_b=0.0190$,
$\omega_c=0.141$, $\Theta_S=1.037$, $\tau=0.0403$, $\log[10^{10}
A_s]$ (0.05Mpc$^{-1}$)$=3.005$, $n_s$(0.05Mpc$^{-1}$)$=0.747$ and
$\alpha_s$(0.05Mpc$^{-1}$)$=-0.150$, respectively. The values of the
minimum $\chi_L^2$ will be presented later in our Table 5. We can find
that the main differences of the two results lie on scales $ \ell \le
20$. TE is relatively not as good a probe as TT, as can also be seen
from the small-scale WMAP3 data.

For the simple PLCDM and RLCDM models some of the resulting
cosmological parameters from WMAP3 alone are somewhat different from
the previously believed concordance cosmology, namely the deviation
from the scale invariant primordial spectrum and a lower $\sigma_8$,
as well as a mild preference of the running $\alpha_s$. For this
purpose we will investigate the possibility how such discrepancies
are affected if we omit low $\ell$  contributions which suffer
from relatively large cosmic variance.
In this section no priors are adopted on $H_0$, $\Omega_b h^2$ and
cosmic age.

\begin{table*}
TABLE 1. Median values and 1$\sigma$ constrains on Power Law
$\Lambda$CDM Model with WMAP1 and WMAP3 and with/without $\ell \le 23$
CMB contributions, shown together with the minimum $\chi_L^2$ values.
``Normal'' stands for the cases where the full range of the WMAP
observational data has been adopted. NO priors have been adopted.
\begin{center}
\begin{tabular}{|c|cc|cc|}

  \hline
&\multicolumn{2}{c|}{WMAP1} &\multicolumn{2}{c|}{WMAP3}\\

No Prior &Normal &$\ell \le 23$ dropped &Normal
&$\ell \le 23$ dropped\\

\hline

$\Omega_b h^2$ &$0.0246^{+0.0026}_{-0.0021}$ &$0.0250\pm0.0025$
&$0.0222\pm0.0007$
&$0.0223\pm0.0010$\\

$\Omega_c h^2$     &$0.115\pm0.018$ &$0.103^{+0.023}_{-0.022}$
&$0.106\pm0.008$ &$0.104\pm0.010$ \\

$\tau$             &$<0.449(95\%)$     &$<0.488(95\%)$
&$<0.139(95\%)$     &$<0.293(95\%)$  \\

$n_s$               &$1.02^{+0.08}_{-0.06}$ &$1.04\pm0.08$
&$0.955\pm0.016$
&$0.960^{+0.033}_{-0.031}$  \\



$\sigma_8$          &$0.949^{+0.143}_{-0.138}$
&$0.939^{+0.132}_{-0.127}$ &$0.762\pm0.050$
&$0.795\pm0.065$   \\

$H_0$               &$75.6^{+9.8}_{-8.1}$ &$80.6^{+12.5}_{-12.1}$
&$73.0\pm3.2$
&$73.6\pm4.9$   \\

  \hline

$\chi_L^2$ &1429.0&1386.0&11252.6&5236.0\\

\hline
\end{tabular}
\end{center}
\end{table*}

First we focus on the case with the WMAP data only. In Table 1 we
list the median values and 1$\sigma$ constrains on PLCDM Model with
WMAP1 and WMAP3 with/without $\ell \le 23$ CMB contributions, shown
together with the minimum $\chi_L^2$ values. ``Normal'' stands for the
cases where the full range of the WMAP observational data has been
adopted. For simplicity the amplitudes of the primordial spectrum
$A_s$ have not been depicted. We should point out that the
likelihood of ``Normal'' WMAP1 is somewhat problematic as mentioned
in the previous parts of the present paper. On the other hand we are
merely using it for comparison. From Table 1 we can find the
improvement of precision on the determination of the cosmological
parameters between WMAP1 and WMAP3. For the parameter $\tau$ only
the $2\sigma$ upper bounds have been delineated. We should point out
the reason that our WMAP1 normal results differ significantly from
that by WMAP team in Ref. \cite{Spergel:2003cb} is mainly due to the
different priors on $\tau$: in our case $\tau$ is almost free while
in Ref. \cite{Spergel:2003cb} a stringent prior $\tau<0.3$ has been
adopted. We can find dramatically for WMAP1 the constraint on $\tau$
changes very little without the presence of low $\ell$ CMB data. This
can be easily understood given the fact that for WMAP1 the
polarization data of EE has not been publicly available. On the
other hand for WMAP3, all of the constraints have been weaker
without the presence of low $\ell$ WMAP and especially for the cases of
$\tau$ and $\sigma_8$. For WMAP3 the eminent effects of omitting $\ell$
$<$ 24 components are that $n_s$ is now consistent with the
Harrison-Zel'dovich-Peebles scale-invariant spectrum and that
the value of
$\sigma_8$ is now nontrivially higher and less stringently
constrained.

\begin{figure}[htbp]
\begin{center}
\includegraphics[scale=0.5]{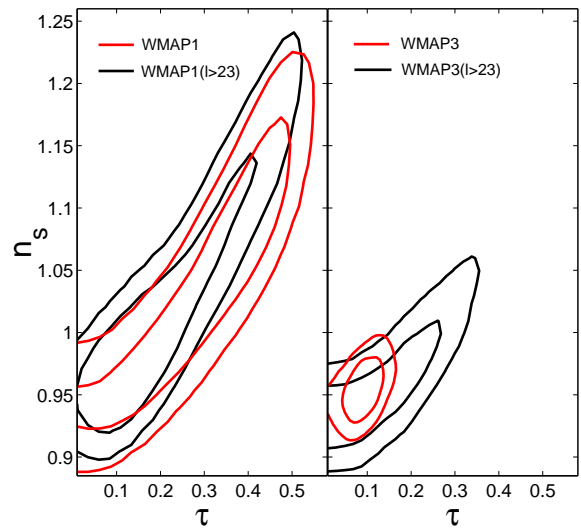}
\caption{ Two dimensional posterior constraints on $n_s$-$\tau$
contours for Power Law $\Lambda$CDM Model with WMAP1 and WMAP3 and
with/without $\ell \le 23$ CMB contributions. The solid lines stand for
1- and 2 $\sigma$ respectively. NO priors have been adopted.
\label{fig:fig001}}
\end{center}
\end{figure}

Compared with WMAP1, in WMAP3 the deviation from scale-invariance is
remarkable due to better measurements of the polarization  and
also more sensitive measurements of the high $\ell$ TT data.
 We should point out our result in Table 1 indicates
that at least for the relevant scales $24\le \ell \le 1000$, WMAP3 is
consistent with the scale invariant spectrum at close to 1$\sigma$.
Thus one may conclude the inconsistency of the scale-invariant
spectrum claimed by WMAP3 analysis is mainly due to the low
multipole components. Our constraint of WMAP3 ``$\ell \le 23$ dropped''
on $\sigma_8$ is in remarkable agreement with the ``$\Lambda$CDM +
scale invariant fluctuations ($n_s=1$)'' case by the WMAP team on
the website \cite{WMAP3IE}, with a considerably larger error bar. On
the other hand, different from our case, the previously reported
``$\Lambda$CDM $+$ primordial spectrum with sharp cutoff and linear
prior'' or `` $\Lambda$CDM $+$ primordial spectrum with sharp cutoff
and log prior''  on the website \cite{WMAP3IE} reported a relatively
stringent constraint on $n_s$ even in the presence of an additional
parameter of the location of the sharp cutoff scale. The reason may
be partially explained by the fact that
the effect of reionization is mainly imprinted on low multipoles of
polarization data, so that $\tau$ cannot be determined precisely
if we omit low $\ell$ components\footnote{ We have recovered the
results ``$\Lambda$CDM $+$
primordial spectrum with sharp cutoff and log prior''  on the
website \cite{WMAP3IE} and also in another case we relaxed the prior
of ``cutoff'', namely we assume ``$\Lambda$CDM $+$ primordial
spectrum with step-like $n_s$ at $k<0.05$ Mpc$^{-1}$ and log prior''.
We find even in this case a nontrivial deviation
from the Harrison-Zel'dovich-Peebles scale-invariant spectrum is
still present: $n_s$ (0.05 Mpc $^{-1}$) $= 0.954 \pm  0.0163$. }.
 As a result the constraint on
$n_s$ is also loosened as seen in Fig.\ \ref{fig:fig001}.


In this figure we delineate the two dimensional posterior
constraints on $n_s$-$\tau$ contours for Power Law $\Lambda$CDM
Model with WMAP1 and WMAP3 and with/without $\ell \le 23$ CMB
contributions, which correspond to the specific subspace of the
likelihood surface in Table 1. The solid lines stand for 1- and 2
$\sigma$ respectively. We can see more clearly that for WMAP1 the
presence of low $\ell$ CMB contributions are almost negligible but
 for WMAP3 the contours are significantly
different, which can be easily understood from the more precise
observations of WMAP3 and the $n_s - \tau$ degeneracy.

\begin{figure}[htbp]
\begin{center}
\includegraphics[scale=0.5]{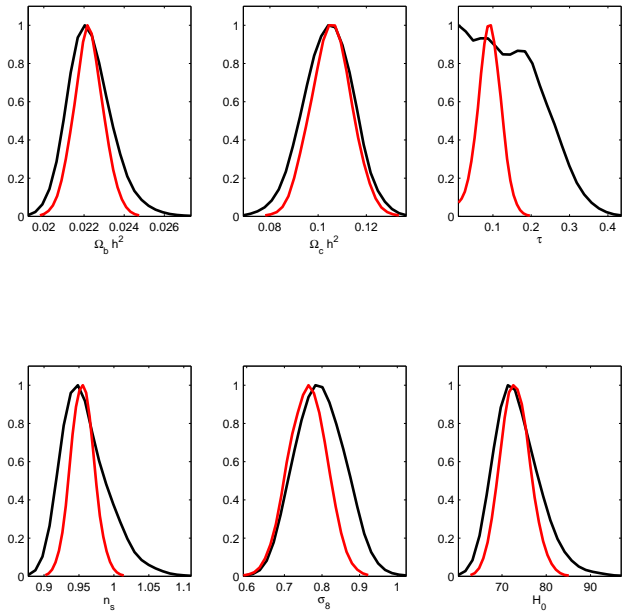}
\caption{ One dimensional posterior constraints on the relevant
cosmological parameters for Power Law $\Lambda$CDM Model with  WMAP3
data, with (red) / without (black) $\ell \le 23$ CMB contributions. The
red lines stand for WMAP3 constraints and the black for constraints
with $\ell\ge 24$ WMAP. NO priors have been adopted.
\label{fig:fig002}}
\end{center}
\end{figure}

Correspondingly, in Fig.~\ref{fig:fig002} we show the one
dimensional posterior constraints on the relevant cosmological
parameters for the PLCDM model with WMAP3 data, with/without $\ell \le
23$ CMB contributions. The red lines stand for WMAP3 constraints and
the black for constraints with $\ell\ge 24$ WMAP3. For the sake of
clarity we have not depicted the WMAP1 cases. One can easily find
the difference brought forth by the low $\ell$ WMAP3 temperature and
polarization contributions, especially for the constraints on
$\tau$, $n_s$ and $\sigma_8$.

\begin{table*}
TABLE 2. Median values and 1$\sigma$ constrains on Power Law
$\Lambda$CDM Model with 2dF, SDSS, SNIa (Riess/SNLS sample) and
WMAP3 (with/without $\ell \le 23$ CMB contributions), shown together
with the minimum $\chi_L^2$ values. ``Normal'' stands for the cases
where the full range of the WMAP3 observational data has been
adopted. NO priors have been adopted.
\begin{center}
\begin{tabular}{|c|cc|cc|}

  \hline
&\multicolumn{2}{c|}{WMAP3+LSS+SNIa(Riess)} &\multicolumn{2}{c|}{WMAP3+LSS+SNIa(SNLS)}\\

No Prior &Normal &$\ell \le 23$ dropped &Normal
&$\ell \le 23$ dropped\\

\hline

$\Omega_b h^2$ &$0.0222\pm0.0007$ &$0.0218\pm0.0007$
&$0.0223\pm0.0007$
&$0.0219\pm0.0007$\\

$\Omega_c h^2$     &$0.113\pm0.005$ &$0.113\pm0.005$
&$0.111\pm0.004$ &$0.111\pm0.004$ \\

$\tau$             &$<0.122(95\%)$     &$<0.167(95\%)$
&$<0.128(95\%)$     &$<0.191(95\%)$  \\

$n_s$               &$0.951\pm0.015$ &$0.937^{+0.017}_{-0.018}$
&$0.953\pm0.016$
&$0.942\pm0.019$  \\



$\sigma_8$          &$0.794\pm0.035$ &$0.788^{+0.047}_{-0.046}$
&$0.786\pm0.035$
&$0.786^{+0.054}_{-0.052}$   \\

$H_0$               &$69.8\pm1.8$ &$68.9\pm1.8$ &$71.0\pm1.8$
&$70.1\pm1.8$   \\

  \hline

$\chi_L^2$ &11495.2&5476.8&11428.4&5410.8\\

\hline
\end{tabular}
\end{center}
\end{table*}

\subsection{WMAP3 data combined with LSS and SNIa}

We now turn to the case  LSS and SNIa are combined with CMB.
 Here we incorporate both
2dF \cite{Cole:2005sx} and SDSS \cite{Tegmark:2003uf} as for LSS data,
 while for SNIa
due to the forementioned reason we combine in one case with the
Riess ``gold'' sample \cite{Riess:2004nr} only and in the other case
with the first year SNLS \cite{snls} only. First we study the PLCDM
model. In Table 2 we show the median values and 1$\sigma$ constrains
on the PLCDM model with 2dF, SDSS, SNIa (Riess/SNLS sample) and
WMAP3 (with/without $\ell \le 23$ CMB contributions), shown together
with the minimum $\chi_L^2$ values. We can easily find that with the
increased observational input, low $\ell$ CMB components lose
 weight for the less relevant parameters like $\Omega_b h^2$,
$\Omega_c h^2$ and $H_0$. On the other hand the parameters like
$\tau$, $n_s$ and $\sigma_8$ are nontrivially affected
depending on the presence of
 $\ell \le 23$ CMB contributions, because low multipole
components are much affected by $\tau$, while the sensitivities of
$n_s$ and especially $\sigma_8$ on the low-$\ell$ components are merely
due to the fact that we are trying to fit the global shape of the
power spectrum with less degrees of freedom. Compared with WMAP3
only case in Table 1 the error bars are much smaller in cases of
``$\ell \le 23$ dropped'' in Table 2. Very intriguingly this leads to
the fact that now the results are again inconsistent with the
Harrison-Zel'dovich-Peebles scale-invariant spectrum, and in cases
one drops the $\ell \le 23$ CMB contributions $n_s$ is even redder
(smaller than 1) than the ``Normal'' case. Also for the case of
WMAP3+LSS+SNIa(Riess), a smaller mean central value of $\sigma_8$ is
present without $\ell \le 23$ CMB contributions, which is in the
opposite direction compared with WMAP only case.
In the left and right columns of Table 2 the
determinations on the cosmological parameters are somewhat different
for the case with WMAP3+LSS+SNIa(Riess) and that with
WMAP3+LSS+SNIa(SNLS), but strikingly consistent at 1$\sigma$. We
will address this issue in more details in the later part of the
current paper.

\begin{figure}[htbp]
\begin{center}
\includegraphics[scale=0.5]{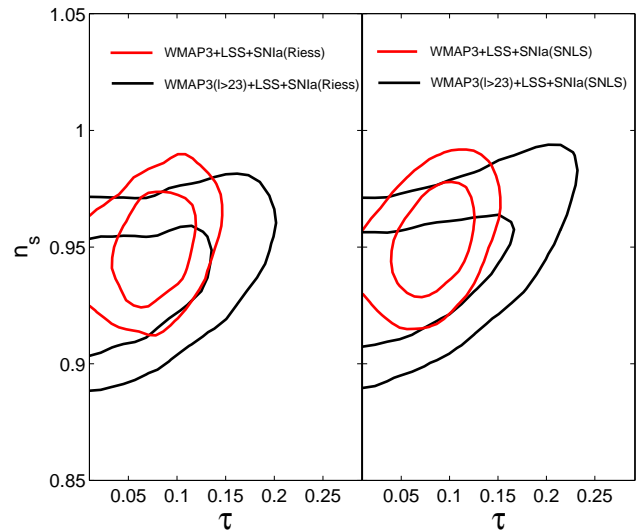}
\caption{ Two dimensional posterior constraints on $n_s$-$\tau$
contours for Power Law $\Lambda$CDM Model with 2dF, SDSS, SNIa
(Riess/SNLS sample) and WMAP3 (with/without $\ell \le 23$ CMB
contributions). The solid lines stand for 1- and 2 $\sigma$
respectively. NO priors have been adopted. \label{fig:fig003}}
\end{center}
\end{figure}

\begin{figure}[htbp]
\begin{center}
\includegraphics[scale=0.5]{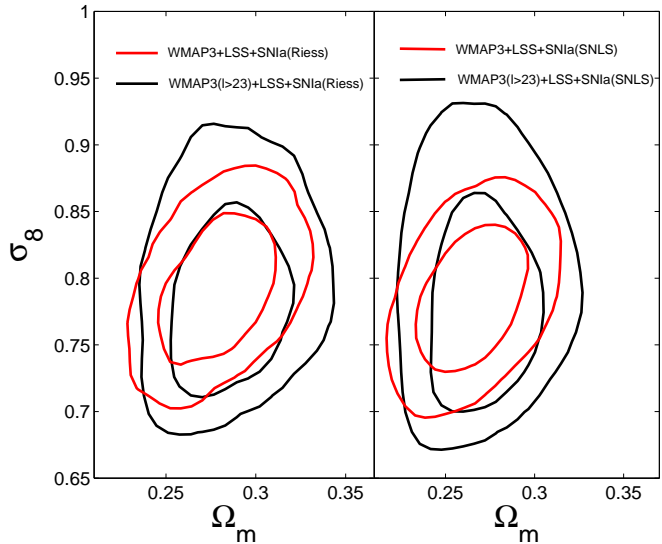}
\caption{ Two dimensional posterior constraints on
$\sigma_8$-$\Omega_m$ contours for Power Law $\Lambda$CDM Model with
2dF, SDSS, SNIa (Riess/SNLS sample) and WMAP3 (with/without $\ell \le
23$ CMB contributions). The solid lines stand for 1- and 2 $\sigma$
respectively. NO priors have been adopted. \label{fig:fig004}}
\end{center}
\end{figure}

In Fig.~\ref{fig:fig003} and Fig.~\ref{fig:fig004} we delineate
the corresponding two dimensional posterior constraints on
$n_s$-$\tau$ contours and $\sigma_8$-$\Omega_m$ contours. The solid
lines stand for 1- and 2 $\sigma$ respectively. The left panels
stand for cases with WMAP3+LSS+SNIa(Riess) with/without $\ell \le 23$
CMB contributions. The right panels show the corresponding cases
with WMAP3+LSS+SNIa(SNLS). We can find that the contours in left
panels are consistent but somewhat different from the right panels,
and 
that a deviation from scale invariance is preferred. The constraints
on $\sigma_8$ are less stringent and more consistent, however not as
significant as the WMAP3 only case as shown in Table 1.

\begin{table*}
TABLE 3. Median values and 1$\sigma$ constrains on Running Spectral
Index $\Lambda$CDM Model with 2dF, SDSS, SNIa (Riess/SNLS sample)
and WMAP3 (with/without $\ell \le 23$ CMB contributions), shown
together with the minimum $\chi_L^2$ values. ``Normal'' stands for the
cases where the full range of the WMAP3 observational data has been
adopted. NO priors have been adopted.
\begin{center}
\begin{tabular}{|c|cc|cc|}

  \hline
&\multicolumn{2}{c|}{WMAP3+LSS+SNIa(Riess)} &\multicolumn{2}{c|}{WMAP3+LSS+SNIa(SNLS)}\\

No Prior &Normal &$\ell \le 23$ dropped &Normal
&$\ell \le 23$ dropped\\

\hline

$\Omega_b h^2$ &$0.0210\pm0.0009$ &$0.0208\pm0.0010$
&$0.0213\pm0.0009$
&$0.0213\pm0.0010$\\

$\Omega_c h^2$     &$0.116\pm0.005$ &$0.118\pm0.005$
&$0.113^{+0.005}_{-0.004}$ &$0.114\pm0.005$ \\

$\tau$             &$<0.148(95\%)$     &$<0.248(95\%)$
&$<0.150(95\%)$     &$<0.263(95\%)$  \\

$n_s$               &$0.880^{+0.038}_{-0.037}$ &$0.862\pm0.049$
&$0.896\pm0.038$
&$0.896\pm0.049$  \\

$\alpha_s$               &$-0.0512^{+0.0251}_{-0.0254}$
&$-0.0717^{+0.0425}_{-0.0419}$ &$-0.0422^{+0.0257}_{-0.0258}$
&$-0.0484\pm0.0441$   \\


$\sigma_8$          &$0.788^{+0.036}_{-0.035}$
&$0.805^{+0.068}_{-0.067}$ &$0.782\pm0.035$
&$0.807^{+0.070}_{-0.068}$   \\

$H_0$               &$67.2\pm2.1$ &$66.5\pm2.3$
&$68.9^{+2.1}_{-2.0}$
&$68.8\pm2.3$   \\

  \hline

$\chi_L^2$ &11491.4&5475.2&11425.8&5410.2\\

\hline
\end{tabular}
\end{center}
\end{table*}

For the cosmological implications of WMAP3+LSS+SNIa now we turn to
the RLCDM model. In Table 3 we delineate the median values and
1$\sigma$ constrains on Running Spectral Index $\Lambda$CDM Model
with 2dF, SDSS, SNIa (Riess/SNLS sample) and WMAP3 (with/without $\ell
\le 23$ CMB contributions), shown together with the minimum $\chi_L^2$
values. Compared with Table 2 we can find that almost all of the
parameters are less stringently constrained with the inclusion of
$\alpha_s$, especially on the relevant ones like $\tau$, $n_s$, and
$\sigma_8$. Due to the inclusion of $\alpha_s$ now all of the
behaviors on the cosmological implications are again in the same
direction as the WMAP3 only case in Table 1, that is, almost
 all of the error
bars are smaller without the presence of $\ell \le 23$ WMAP3
contributions, and a larger mean center value of $\sigma_8$ is
present when dropping small $\ell$ CMB contributions. On the constraint
on $\alpha_s$ itself we can find from Table 3 the case of ``Normal''
WMAP3+LSS+SNIa(Riess) favors a negative running of $\alpha_s$ to nearly
$2 \sigma$. For the other three cases listed a running is
less favored, but still present and in some sense the preference of
the negative running is fairly independent of low $\ell$ WMAP3
contributions, which will be explored in more details in the
remaining part of this paper.

\begin{figure}[htbp]
\begin{center}
\includegraphics[scale=0.5]{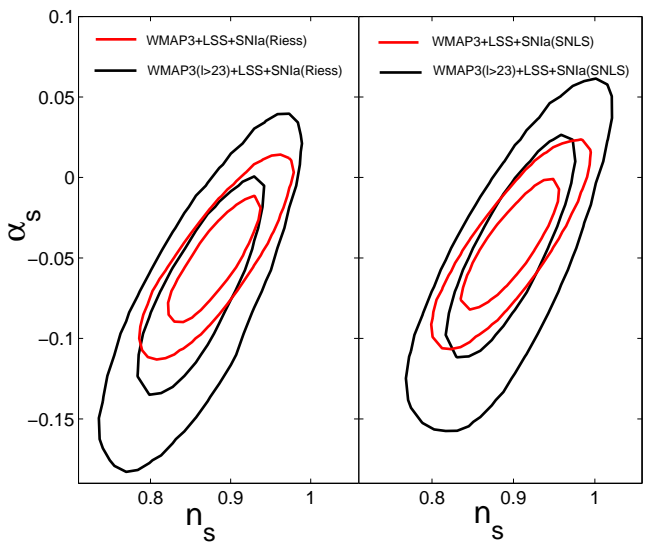}
\caption{ Two dimensional posterior constraints on $n_s$-$\alpha$
contours for Running Spectral Index $\Lambda$CDM Model with 2dF,
SDSS, SNIa (Riess/SNLS sample) and WMAP3 (with/without $\ell \le 23$
CMB contributions). The solid lines stand for 1- and 2 $\sigma$
respectively. NO priors have been adopted. \label{fig:fig005}}
\end{center}
\end{figure}

In Fig.~\ref{fig:fig005} we show the corresponding two dimensional
posterior constraints on $n_s$-$\alpha$ contours for Running
Spectral Index $\Lambda$CDM Model with 2dF, SDSS, SNIa (Riess/SNLS
sample) and WMAP3 (with/without $\ell \le 23$ CMB contributions). The
solid lines stand for 1- and 2 $\sigma$ respectively. We can find
that in the case of dropping small $\ell$ CMB contributions the
discrepancy between the left and the right panels are nontrivial and
this needs to be reconsidered in more details with the accumulations
of the currently ongoing SNIa projects. Also the
Harrison-Zel'dovich-Peebles scale-invariant spectrum lies within the
2$\sigma$ range for the case of WMAP3+LSS+SNIa(SNLS) without small
$\ell$ CMB contributions, which is different from the other three
cases. It is also intriguing  that
 in the case of ``All'' combination of  WMAP team a negative
running is preferred to more than 2$\sigma$, namely,
 $\alpha_s = -0.061\pm
0.023$ \cite{WMAP3IE}. Their case of ``All'' differs from our
analysis here regarding the following aspects: they have included
small scale CMB observations and combined the SNIa data of Riess and
SNLS samples simultaneously, made some detailed analysis on SDSS and
2dF bias factors and they have considered the secondary effects of
SZ on WMAP3. And it is interesting to imagine the consequences
without small $\ell$ CMB contributions for their case of ``All''.

\begin{table*}
TABLE 4. Median values and 1$\sigma$ constrains on Power Law
$\Lambda$CDM Model with WMAP3 (with/without $\ell \le 23$ CMB
contributions) combined with LUQAS+CROFT sample of Lyman $\alpha$ as
analyzed in Ref. \cite{Viel:2004bf} (dubbed V-Ly$\alpha$) or the SDSS Lyman
$\alpha$ sample from Ref. \cite{McDonald:2004xn} (dubbed S-Ly$\alpha$),
shown together with the minimum $\chi_L^2$ values. ``Normal'' stands
for the cases where the full range of the WMAP3 observational data
has been adopted. NO priors have been adopted.
\begin{center}
\begin{tabular}{|c|cc|cc|}

  \hline
&\multicolumn{2}{c|}{WMAP3 + V-Ly$\alpha$} &\multicolumn{2}{c|}{WMAP3 + S-Ly$\alpha$}\\

No Prior &Normal &$\ell \le 23$ dropped &Normal
&$\ell \le 23$ dropped\\

\hline

$\Omega_b h^2$ &$0.0223\pm0.0007$ &$0.0223^{+0.0009}_{-0.0010}$
&$0.0228\pm0.0007$
&$0.0230\pm0.0010$\\

$\Omega_c h^2$     &$0.110\pm0.008$ &$0.109\pm0.009$
&$0.120\pm0.006$ &$0.112\pm0.010$ \\

$\tau$             &$<0.139(95\%)$     &$<0.283(95\%)$
&$<0.149(95\%)$     &$<0.313(95\%)$  \\

$n_s$               &$0.955\pm0.017$ &$0.958\pm0.029$
&$0.964\pm0.016$
&$0.977^{+0.032}_{-0.031}$  \\



$\sigma_8$          &$0.787\pm0.047$ &$0.826^{+0.060}_{-0.059}$
&$0.857\pm0.024$
&$0.891\pm0.035$   \\

$H_0$               &$71.6^{+3.0}_{-2.9}$ &$71.8\pm4.4$
&$68.8\pm2.5$
&$71.9^{+5.0}_{-4.9}$   \\

  \hline

$\chi_L^2$ &11280.4&5263.2&11444.6&5425.2\\

\hline
\end{tabular}
\end{center}
\end{table*}

Although  one often takes the risks of
uncontrolled systematics
\cite{Lesgourgues:2006nd}, Ly$\alpha$
can probe scales as
small as $k \sim 1 h$ Mpc $^{-1}$ to yield
the amplitude of density fluctuations $\sigma_8$
directly unlike the other data sets.
  Hence it is urged to probe the
possible discrepancy between WMAP3 and Ly$\alpha$ in the determination of
the cosmological parameters, especially on the value of $\sigma_8$.
In Table 4 we delineate the median values and 1$\sigma$ constrains
on Power Law $\Lambda$CDM Model with WMAP3 (with/without $\ell \le 23$
CMB contributions) combined with V-Ly$\alpha$/ S-Ly$\alpha$, shown together with
the minimum $\chi_L^2$ values. One can find again that compared with
the ``Normal'' cases of WMAP3+Ly$\alpha$ combinations where a deviation
from scale-invariance is still significant and the discrepancies on
the determination of $\sigma_8$ are rather large among WMAP3 alone,
WMAP3+V-Ly$\alpha$ and WMAP3+S-Ly$\alpha$ combinations. In cases of dropping
small $\ell$ WMAP3 contributions, $n_s$ is nontrivially consistent with
scale-invariance.

\begin{figure}[htbp]
\begin{center}
\includegraphics[scale=0.55]{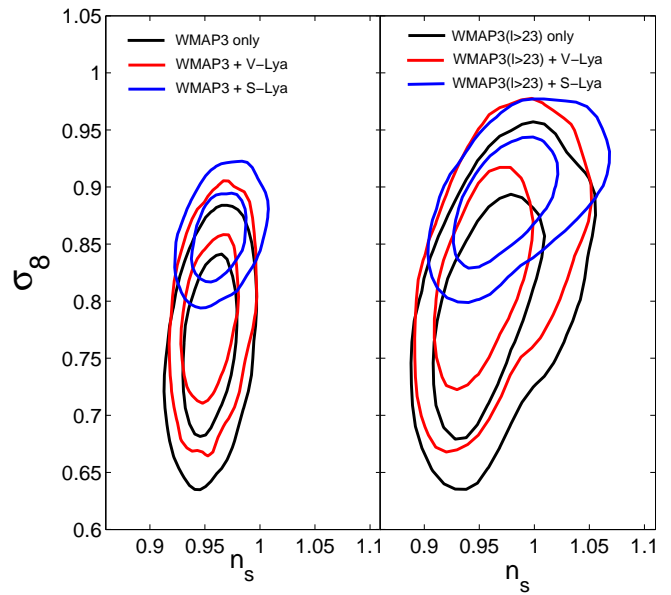}
\caption{ Two dimensional posterior constraints on $n_s$-$\sigma_8$
contours for Power Law $\Lambda$CDM Model with WMAP3 only
(with/without $\ell \le 23$ CMB contributions) or WMAP3 combined with
LUQAS+CROFT sample of Lyman $\alpha$ as analyzed in
Ref. \cite{Viel:2004bf} (dubbed V-Ly$\alpha$) or with the SDSS Lyman
$\alpha$ sample from Ref. \cite{McDonald:2004xn} (dubbed S-Ly$\alpha$). The
solid lines stand for 1- and 2 $\sigma$ respectively. No priors have
been adopted. \label{fig:fig006}}
\end{center}
\end{figure}

In Fig.~ \ref{fig:fig006} we show the corresponding two dimensional
posterior constraints on $n_s$-$\sigma_8$ contours for the PLCDM
model with WMAP3 (with/without $\ell \le 23$ CMB contributions)
combined with  V-Ly$\alpha$/ S-Ly$\alpha$. The solid lines stand for 1- and 2
$\sigma$ respectively. In the left panel we delineate the ``Normal''
cases and in the right panel we show the corresponding cases where
the CMB data are not used for $\ell \le 23$. Our left panel is almost
exactly the same as depicted previously in Ref. \cite{Viel:2006yh},
where the discrepancies on $\sigma_8$ are noteworthy for the listed
three kinds of different data combinations. On the other hand as
shown in our right panel, in cases we neglect the low $\ell$ WMAP3
contributions, all of the contours get enlarged and these three
cases are consistent with each other in a wider range of parameters.

\section{Probing the scale dependence of the primordial spectrum}

\begin{table*}
TABLE 5. Median values and 1$\sigma$ constrains on Running Spectral
Index $\Lambda$CDM Model with WMAP1 and WMAP3 and with/without $\ell
\le 23$ CMB contributions, shown together with the minimum $\chi_L^2$
values. ``Normal'' stands for the cases where the full range of the
WMAP observational data has been adopted. HST, BBN and age priors
have been adopted.
\begin{center}
\begin{tabular}{|c|cc|cc|}

  \hline
&\multicolumn{2}{c|}{WMAP1} &\multicolumn{2}{c|}{WMAP3}\\

 &Normal &$\ell \le 23$ dropped &Normal
&$\ell \le 23$ dropped\\

\hline

$\Omega_b h^2$ &$0.0248^{+0.0028}_{-0.0027}$
&$0.0244^{+0.0029}_{-0.0031}$ &$0.0209\pm0.0010$
&$0.0198^{+0.0014}_{-0.0015}$\\

$\Omega_c h^2$     &$0.0986^{+0.0235}_{-0.0218}$ &$0.106\pm0.025$
&$0.115^{+0.010}_{-0.011}$ &$0.132\pm0.019$ \\

$\tau$             &$<0.564(95\%)$     &$<0.529(95\%)$
&$<0.155(95\%)$     &$<0.274(95\%)$  \\

$n_s$               &$0.985^{+0.081}_{-0.084}$
&$0.996^{+0.121}_{-0.127}$        &$0.876^{+0.046}_{-0.048}$
&$0.789^{+0.088}_{-0.090}$  \\

$\alpha_s$               &$-0.0885^{+0.0474}_{-0.0466}$
&$-0.0432^{+0.0891}_{-0.0886}$ &$-0.0550^{+0.0307}_{-0.0317}$
&$-0.135\pm0.071$   \\


$\sigma_8$          &$0.936^{+0.171}_{-0.131}$
&$0.939^{+0.168}_{-0.134}$ &$0.782^{+0.048}_{-0.049}$
&$0.831\pm0.063$   \\

$H_0$               &$82.9^{+12.4}_{-13.7}$ &$78.6^{+13.7}_{-13.7}$
&$67.7^{+4.5}_{-4.3}$
&$61.4\pm7.0$   \\

  \hline

$\chi_L^2$ &1425.4&1386.2&11249.6&5232.0\\

\hline
\end{tabular}
\end{center}
\end{table*}

In the above investigations on the discrepancies between
the implications of
the current observations with and
without low $\ell$ WMAP3 contributions we
have adopted no priors on $H_0$, $\Omega_b h^2$ and cosmic age. In
the following study we mainly focus on the cosmological implications
on the scale dependence of the primordial spectrum, where we prefer
to include the physical priors from HST, BBN and cosmic age. A
running of the primordial spectrum will be studied extensively for
our remaining studies.

\subsection{WMAP data alone}

We also start from the WMAP only case. In Table 5 we list the median
values and 1$\sigma$ constrains on Running Spectral Index
$\Lambda$CDM Model with WMAP1 and WMAP3 and with/without $\ell \le 23$
CMB contributions, shown together with the minimum $\chi_L^2$ values.
As shown explicitly for WMAP1 while a negative running of $\alpha_s$
is close to 2$\sigma$ for the ``Normal'' case, no running is
in full consistency with the case  ``$\ell \le 23$ dropped''. Our
result is consistent with the analysis by
Bridle et al.\  \cite{Bridle:2003sa}. On the other hand, given the
problematic likelihood of WMAP1 on low $\ell$ components,
our analysis with ``$\ell \le 23$
dropped'' should be more conservative but also more robust compared
with the analysis in Ref. \cite{Bridle:2003sa}. With the accumulation
of the observations and better understanding on the systematics,
the more precise WMAP3 shows a very different behavior on the shape
of the primordial spectrum, that is, a negative running is preferred to
nearly 2$\sigma$ for both of the cases with/without
small $\ell$ CMB contributions. Also the parameter space for $\sigma_8$
is more consistent with the previously well-accepted values for the
analysis of the cosmic structure formations.

\begin{figure}[htbp]
\begin{center}
\includegraphics[scale=0.5]{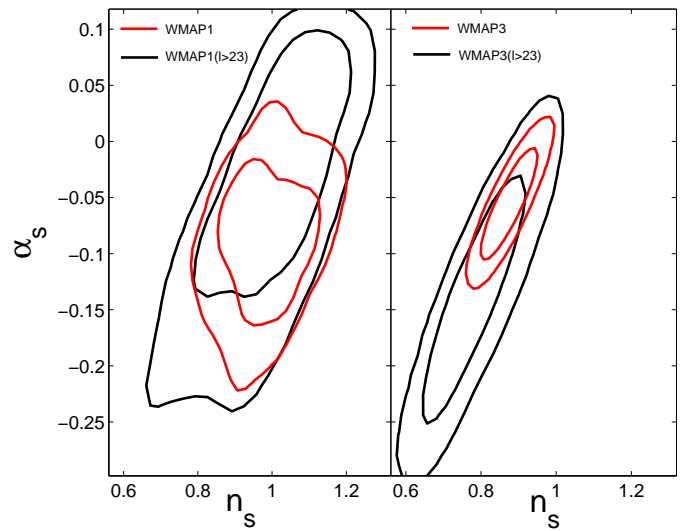}
\caption{ Two dimensional posterior constraints on $n_s$-$\alpha_s$
contours for Running Spectral Index $\Lambda$CDM Model with WMAP1
and WMAP3 and with/without $\ell \le 23$ CMB contributions. The solid
lines stand for 1- and 2 $\sigma$ respectively. HST, BBN and age
priors have been adopted. \label{fig:fig007}}
\end{center}
\end{figure}

Correspondingly in Fig.~\ref{fig:fig007} we show the resulting two
dimensional posterior constraints on $n_s$-$\alpha_s$ contours for
the RLCDM model with WMAP1 and WMAP3 and with/without $\ell \le 23$ CMB
contributions. The solid lines stand for 1- and 2 $\sigma$
respectively. While for WMAP1 even a large positive running is
consistent when neglecting the small $\ell$ CMB contributions, for
WMAP3 we can easily find the significantly different likelihood
space where a large negative running takes the main parameter space.
And also for ``Normal'' WMAP3 a negative running is nontrivially
preferred with a higher precision compared with WMAP1, where the
actual preference should be diminished if the correct likelihood
functions is adopted \cite{Slosar:2004fr}.

\begin{table*}
TABLE 6. Median values and 1$\sigma$ constrains on Running Spectral
Index $\Lambda$CDM Model with 2dF, SDSS and WMAP3 (with/without $\ell
\le 23$ CMB contributions), shown together with the minimum $\chi_L^2$
values. ``Normal'' stands for the cases where the full range of the
WMAP3 observational data has been adopted. HST, BBN and age priors
have been adopted.
\begin{center}
\begin{tabular}{|c|cc|cc|cc|}

  \hline
&\multicolumn{2}{c|}{WMAP3 + SDSS} &\multicolumn{2}{c|}{WMAP3 + 2dF} &\multicolumn{2}{c|}{WMAP3 + SDSS + 2dF}\\

 &Normal &$\ell \le 23$ dropped  &Normal &$\ell \le 23$ dropped &Normal
&$\ell \le 23$ dropped\\

\hline

$\Omega_b h^2$
&$0.0207^{+0.0010}_{-0.0009}$&$0.0194^{+0.0012}_{-0.0011}$&$0.0213\pm0.0009$&$0.0214\pm0.0010$
&$0.0210\pm0.0010 $ &$0.0205\pm0.0012$\\

$\Omega_c h^2$  &$0.123\pm0.008$&$0.137\pm0.012$
&$0.111\pm0.006$&$0.112\pm0.007$
&$0.116\pm0.006$       &$0.121\pm0.009$  \\

$\tau$      &$<0.147(95\%)$& $<0.226(95\%)$
&$<0.153(95\%)$&$<0.286(95\%)$
&$<0.149(95\%)$        &$<0.254(95\%)$   \\

$n_s$      &$0.860^{+0.043}_{-0.042}$&  $0.767^{+0.067}_{-0.068}$
&$0.899\pm0.038$&$0.892^{+0.050}_{-0.052}$
 &$0.882^{+0.041}_{-0.042}$        &$0.845^{+0.058}_{-0.069}$   \\

$\alpha_s$     &$-0.0655^{+0.0280}_{-0.0279}$&
$-0.148^{+0.056}_{-0.059}$
&$-0.0410^{+0.0256}_{-0.0258}$&$-0.0570^{+0.0467}_{-0.0477}$
  &$-0.0505^{+0.0274}_{-0.0271}$        &$-0.0855^{+0.0550}_{-0.0551}$   \\


$\sigma_8$    &$0.812\pm0.040$&   $0.833^{+0.061}_{-0.060}$
&$0.772^{+0.038}_{-0.036}$&$0.812\pm0.073$
  &$0.788\pm0.036$         &$0.808^{+0.070}_{-0.065}$   \\

$H_0$     &$64.4\pm3.2$&  $59.1^{+4.2}_{-4.3}$
&$69.8\pm2.5$&$69.6\pm3.0$
   &$67.5^{+2.8}_{-2.7}$         &$65.4^{+3.9}_{-4.0}$   \\

  \hline

$\chi_L^2$ &11267.0&5247.6&11288.8&5273.8&11307.8&5291.8\\

\hline
\end{tabular}
\end{center}
\end{table*}

\subsection{Combination with other datasets}

For a next step we consider the implications of WMAP3 in combination
with LSS in great details. In Table 6 we show the median values and
1$\sigma$ constrains on the RLCDM model with 2dF, SDSS and WMAP3
(with/without $\ell \le 23$ CMB contributions), shown together with the
minimum $\chi_L^2$ values. The separate combinations of WMAP with SDSS
and 2dF have also been included for the detailed investigations. The
discrepancy between SDSS and 2dF is most significantly depicted in
the measurement of the matter density $\Omega_c h^2$, where the
tension is even more eminent in cases without low $\ell$ WMAP3
contributions. Our analysis here is consistent with the results by
the WMAP group \cite{WMAP3IE}. While WMAP3's better measurements of
the TT third peak help a lot on the determination of the matter
density, in the concordance cosmology the measurements of $\Omega_c
h^2$ from LSS are also nontrivial. The next release of SDSS power
spectrum is expected to appear soon with much higher
precision\footnote{Max Tegmark, private communications.}, the
discrepancy on the matter density is hopefully to be verified or
eliminated. Given the current LSS data we find that for the
combination of WMAP3+SDSS the preference of running is larger than
2$\sigma$ regardless of the presence of low $\ell$ CMB. While for
the WMAP3+2dF combination or that with WMAP3+2dF+SDSS, preference of
a negative
running is less prominent, which nevertheless does exist.

\begin{figure}[htbp]
\begin{center}
\includegraphics[scale=0.5]{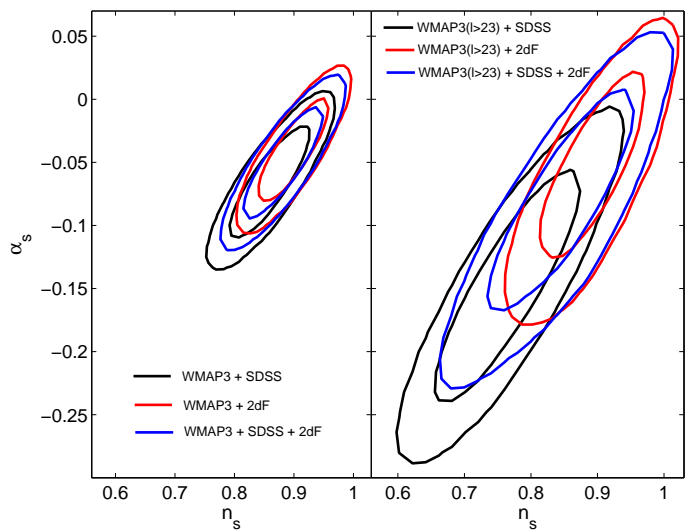}
\caption{ Two dimensional posterior constraints on $n_s$-$\alpha$
contours for Running Spectral Index $\Lambda$CDM Model with 2dF,
SDSS and WMAP3 (with/without $\ell \le 23$ CMB contributions). The
solid lines stand for 1- and 2 $\sigma$ respectively. HST, BBN and
age priors have been adopted. \label{fig:fig008}}
\end{center}
\end{figure}

\begin{figure}[htbp]
\begin{center}
\includegraphics[scale=0.5]{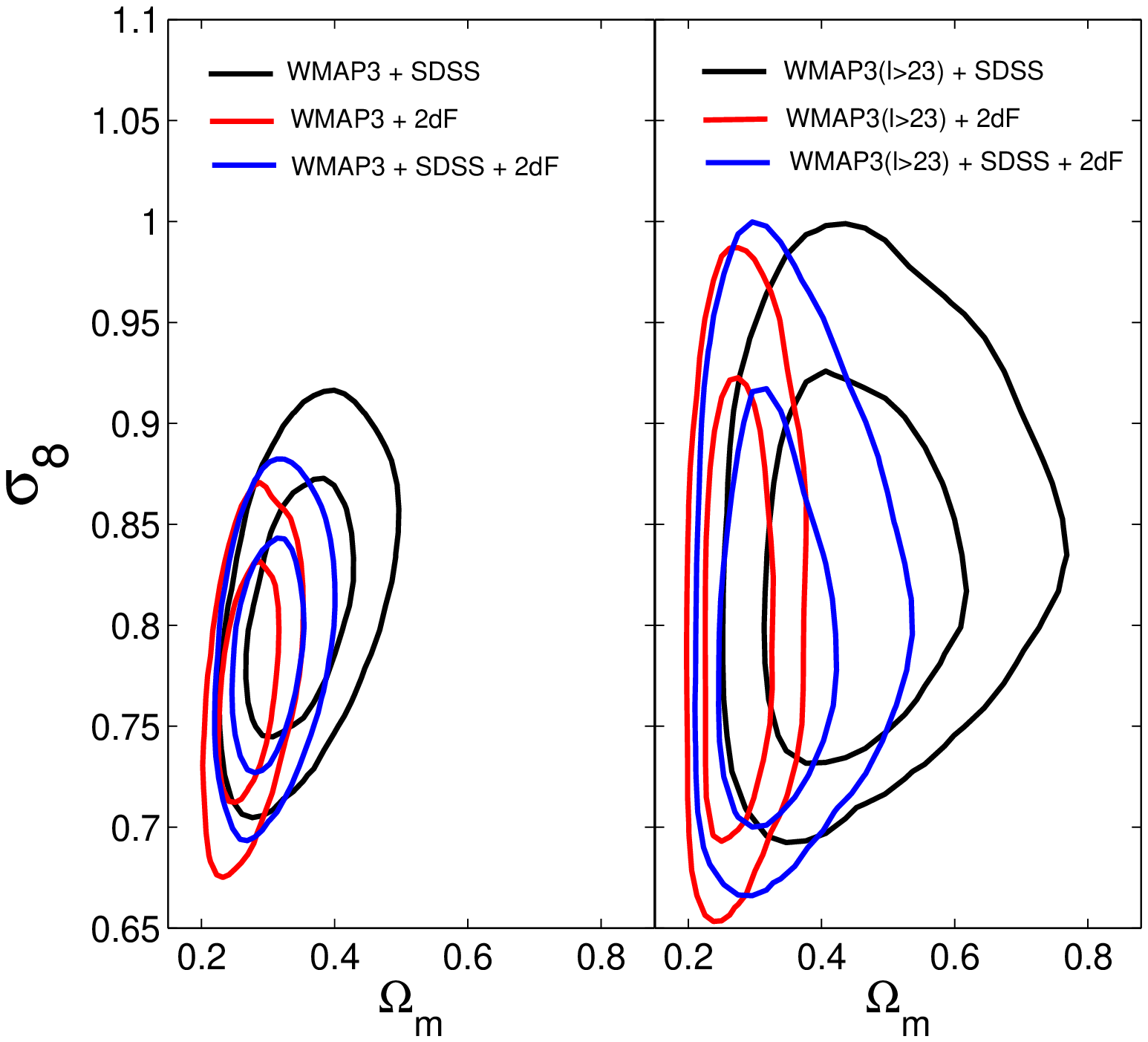}
\caption{ Two dimensional posterior constraints on
$\Omega_m$-$\sigma_8$ contours for Running Spectral Index
$\Lambda$CDM Model with 2dF, SDSS and WMAP3 (with/without $\ell \le 23$
CMB contributions). The solid lines stand for 1- and 2 $\sigma$
respectively. HST, BBN and age priors have been adopted.
\label{fig:fig009}}
\end{center}
\end{figure}

In Fig.~\ref{fig:fig008} and Fig.~\ref{fig:fig009} we depict the
corresponding two dimensional posterior constraints on the
$n_s$-$\alpha$ contours and on the $\Omega_m$-$\sigma_8$ contours,
respectively. The solid lines stand for 1- and 2 $\sigma$
respectively. The left panels show the ``Normal'' constraints and
the right ones show the cases where low $\ell$ WMAP3 are not used for
the global fittings.


\begin{table*}
TABLE 7. Median values and 1$\sigma$ constrains on Power Law,
Running Spectral Index $\Lambda$CDM Model and Running Spectral Index
$\Lambda$CDM Model with step-like $n_s$ truncated at $k= 0.1$
Mpc$^{-1}$(SRLCDM), as measured by  WMAP3 in combination with
LUQAS+CROFT sample of Lyman $\alpha$ as analyzed in
Ref. \cite{Viel:2004bf} (dubbed V-Ly$\alpha$) or the SDSS Lyman $\alpha$
sample from Ref. \cite{McDonald:2004xn} (dubbed S-Ly$\alpha$), shown
together with the minimum $\chi_L^2$ values. HST, BBN and age priors
have been adopted.
\begin{center}
\begin{tabular}{|c|ccc|cc|ccc|}

  \hline
&\multicolumn{3}{c|}{PLCDM} &\multicolumn{2}{c|}{RLCDM} &\multicolumn{3}{c|}{SRLCDM}\\

 &WMAP3 only &V-Ly$\alpha$ &S-Ly$\alpha$ &V-Ly$\alpha$ &S-Ly$\alpha$ &WMAP3 only &V-Ly$\alpha$ &S-Ly$\alpha$ \\

\hline

$10^2\Omega_b h^2$ &$2.22\pm0.07$
&$2.24\pm0.07$&$2.27\pm0.06$&$2.19\pm0.07$ &$2.25\pm0.06$
&$2.12\pm0.09$ &$2.12\pm0.09$
&$2.12\pm0.09$  \\

$\Omega_c h^2$  &$0.106^{+0.007}_{-0.008}$ &$0.110\pm0.007$&
$0.120\pm0.005$ &$0.115\pm0.008$ &$0.122\pm0.006$ &$0.112\pm0.009$
&$0.118\pm0.009$ &$0.125\pm0.006$   \\

$\tau$   &$<0.136(95\%)$     &$<0.138(95\%)$&   $<0.147(95\%)$
&$<0.152(95\%)$ &$<0.157(95\%)$ &$<0.157(95\%)$
&$<0.162(95\%)$     &$<0.170(95\%)$     \\

$n_s$   &$0.953\pm0.016$   &$0.956\pm0.015$& $0.964\pm0.015$
&$0.923^{+0.026}_{-0.027}$ &$0.946\pm0.020$ &$0.887\pm0.042$
&$0.882^{+0.039}_{-0.040}$
&$0.875^{+0.039}_{-0.038}$    \\

$n_s'$   &$-$   &$-$ &$-$         &$-$ &$-$ &$No~constraint$
&$0.929\pm0.090$
&$0.958\pm0.063$    \\

$\alpha_s$   &$-$    &$-$ &$-$ &$-0.0268^{+0.0171}_{-0.0176}$
&$-0.0154^{+0.0121}_{-0.0116}$ &$-0.0487^{+0.0284}_{-0.0282}$
&$-0.0552^{+0.0269}_{-0.0272}$
&$-0.0635^{+0.0260}_{-0.0257}$   \\


$\sigma_8$   &$0.760^{+0.046}_{-0.047}$  &$0.779\pm0.045$&
$0.856\pm0.025$ &$0.804^{+0.045}_{-0.047}$ &$0.862\pm0.025$
&$<2.36(95\%)$ &$0.818\pm0.045$
&$0.863^{+0.028}_{-0.027}$      \\

$H_0$   &$72.8\pm3.0$   &$71.6\pm2.8$& $68.9\pm2.3$
&$69.4^{+3.2}_{-3.1}$ &$68.0\pm2.5$ &$69.1\pm3.7$ &$67.3\pm3.5$
&$65.0\pm2.8$    \\

\hline

$\chi_L^2$ &11252.8&11280.0&11444.6&11277.4&11443.4&11249.8&11276.6&11439.6 \\

  \hline
\end{tabular}
\end{center}
\end{table*}

To measure directly the quantity of $\sigma_8$ one needs to include
the Ly$\alpha$ observations. Now we probe the scale dependence of the
primordial spectrum in the light of WMAP3 and WMAP3 in combinations
with the LUQAS+CROFT sample of Lyman $\alpha$ as analyzed in
Ref. \cite{Viel:2004bf} (V-Ly$\alpha$) or the SDSS Lyman $\alpha$ sample
from Ref. \cite{McDonald:2004xn} (S-Ly$\alpha$). We investigate the three
cases: PLCDM, RLCDM and SRLCDM cosmology where for SRLCDM we mean
the Running Spectral Index $\Lambda$CDM Model with step-like $n_s$
truncated at $k= 0.1$ Mpc$^{-1}$, with $n'_s$ being the additionally
introduced parameter which represents the spectral index on scales
$k\ge 0.1$ Mpc$^{-1}$. Note in this case we have assumed a constant
$n'_s$. In Table 7 we list the median values and 1$\sigma$
constrains on the determinations of the relevant cosmological
parameters for the three cases.
For the PLCDM case a deviation from
scale-invariance is present for all of the three cases in the left
column of Table 7. For the RLCDM case as we are assuming a constant
running on all scales, a vanishing running is consistent in either
combination of WMAP3 with V-Ly$\alpha$ or S-Ly$\alpha$.
For the case of SRLCDM we
find many interesting implications. Because WMAP3 cannot probe scales
with $k\ge 0.1$ Mpc$^{-1}$, we cannot get any constraint on the
parameter $n'_s$ from WMAP3 alone.
 And for this case we get $\sigma_8<2.36$ at 95\%
confidence level, which we will explain later in more detail.
Comparing Table 7 with Table 5, we can find that the SRLCDM case of
WMAP3 only is similar to that in Table 5 for the RLCDM case only,
and a running is favored at less than 2$\sigma$. Interestingly for
the cases of WMAP3 in combinations with V-Ly$\alpha$ or S-Ly$\alpha$,
a negative
running is preferred at $>2 \sigma$ on scales $k\le 0.1$ Mpc$^{-1}$
for both of the cases and the implications on $\sigma_8$ turn out to
be more consistent. This should be relevant to the fact that the
introduction of Ly$\alpha$ data changes the preference of $\Omega_m$
compared with WMAP3 only case, where $\Omega_m$ is much relevant to
the TT third peak. A more intriguing aspect comes from the
determination on $n'_s$ for the cases in combinations with V-Ly$\alpha$ or
S-Ly$\alpha$: for both cases the
 Harrison-Zel'dovich-Peebles scale-invariant
spectrum fits well the observations on scales $k\ge 0.1$
Mpc$^{-1}$. On the other hand compared with a constant running RLCDM
case, a constant negative running for $k\le 0.1$ Mpc$^{-1}$ and nearly
scale invariant spectrum for scales $k\ge 0.1$ Mpc$^{-1}$ is
preferred at close to $2\sigma$ for the WMAP3+S-Ly$\alpha$
combination and
similar behavior also exhibits for the WMAP3+V-Ly$\alpha$ combination,
which is relatively weaker than the former case.

\begin{figure*}[htbp]
\begin{center}
\includegraphics[scale=0.6]{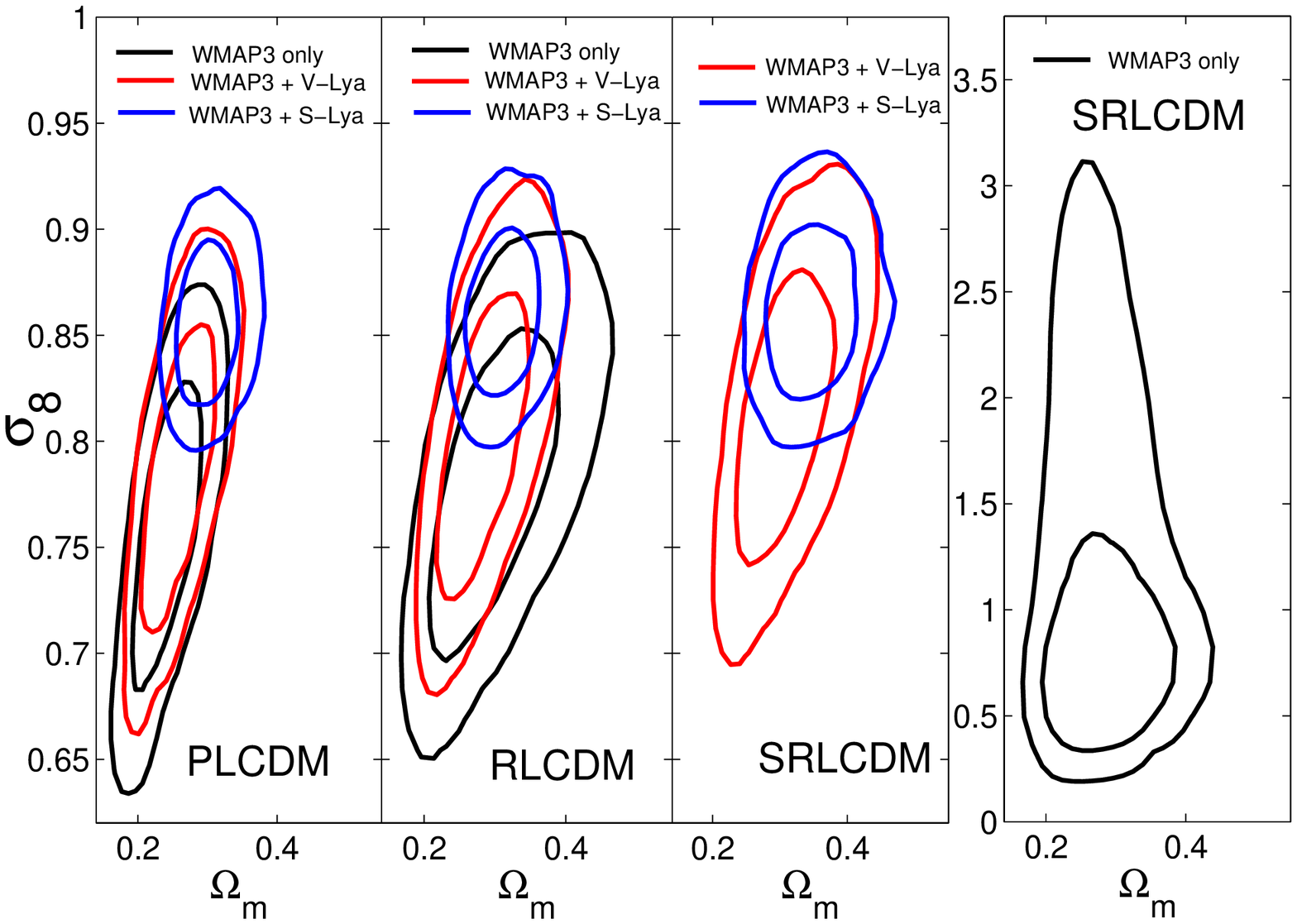}
\caption{Two dimensional posterior constraints on
$\Omega_m$-$\sigma_8$ contours for Power Law, Running Spectral Index
$\Lambda$CDM Model and Running Spectral Index $\Lambda$CDM Model
with step-like $n_s$ truncated at $k= 0.1$ Mpc$^{-1}$
(PLCDM/RLCDM/SRLCDM), as measured by  WMAP3 alone and WMAP3 in
combination with the LUQAS+CROFT sample of Lyman $\alpha$ as
analyzed in Ref. \cite{Viel:2004bf} (dubbed V-Ly$\alpha$) or the
SDSS Lyman $\alpha$ sample from Ref. \cite{McDonald:2004xn} (dubbed
S-Ly$\alpha$). HST, BBN and age priors have been adopted. See the
text for more details. \label{fig:fig0011}}
\end{center}
\end{figure*}

In Fig.~\ref{fig:fig0011} we depict the two dimensional posterior
constraints on $\Omega_m$-$\sigma_8$ contours for Power Law, Running
Spectral Index $\Lambda$CDM Model and Running Spectral Index
$\Lambda$CDM Model with step-like $n_s$ truncated at $k= 0.1$
Mpc$^{-1}$, as measured by  WMAP3 alone and WMAP3 in combination
with the LUQAS+CROFT sample of Lyman $\alpha$ as analyzed in
Ref. \cite{Viel:2004bf} (dubbed V-Ly$\alpha$) or the SDSS Lyman $\alpha$
sample from Ref. \cite{McDonald:2004xn} (dubbed S-Ly$\alpha$).
We can easily
find that the discrepancy  is the largest in PLCDM and as we
introduce the additional parameters like a constant running and also
then work in the SRLCDM framework, they gradually become consistent
with each other. And for the case of SRLCDM, as $n'_s$ is not
constrained, the resulting constraint on $\sigma_8$ would definitely
be extremely weak and we put the WMAP3 constraint alone on the right
panel.

\begin{figure*}[htbp]
\begin{center}
\includegraphics[scale=0.8]{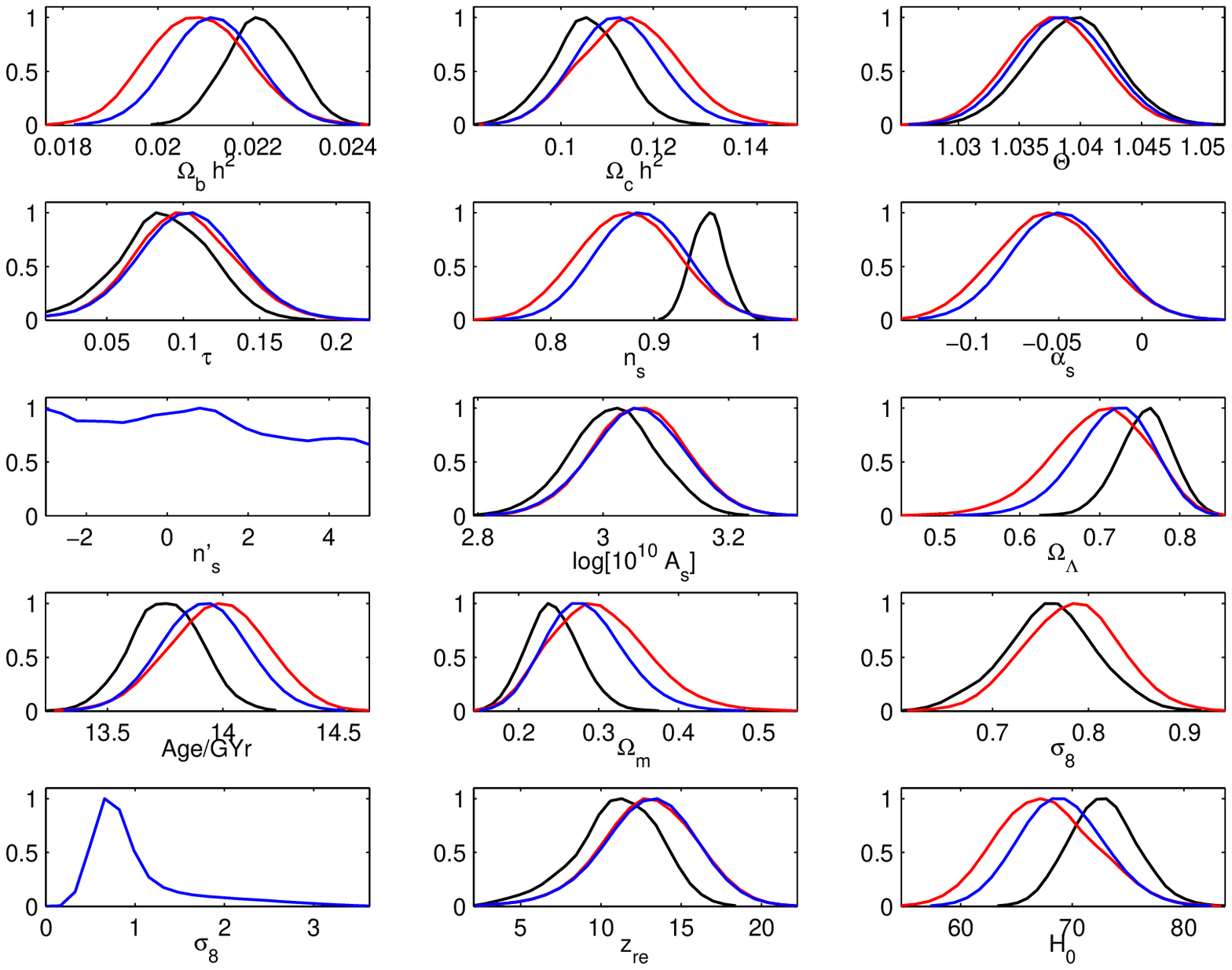}
\caption{One dimensional posterior constraints on the relevant
cosmological parameters for Power Law $\Lambda$CDM Model(PLCDM),
Running Spectral Index $\Lambda$CDM Model(RLCDM) and Running
Spectral Index $\Lambda$CDM Model with step-like $n_s$ truncated at
$k= 0.1$ Mpc$^{-1}$(SRLCDM), as measured by WMAP3 alone. The black
lines stand for PLCDM, red lines for RLCDM and and the blue for
SRLCDM model. HST, BBN and age priors have been adopted.
\label{fig:wmapx}}
\end{center}
\end{figure*}

For a better study on the cosmological constraint by WMAP3 alone, in
Fig.~\ref{fig:wmapx} we show the one dimensional posterior
constraints on the relevant cosmological parameters for the PLCDM
model, the RLCDM model and  the SRLCDM model respectively, as
measured by WMAP3 alone. The black lines stand for PLCDM, red lines
for RLCDM and  the blue for SRLCDM model. We should point out
that although our chains have converged, our results on the SRLCDM
model is much dependent on the prior adopted on the parameter
$n'_s$, which is easily understood given the fact that $n'_s$ is
unconstrained for WMAP3 alone. And a weaker prior on $n'_s$ would
lead to a weaker constraint on $\sigma_8$.

\begin{figure*}[htbp]
\begin{center}
\includegraphics[scale=1]{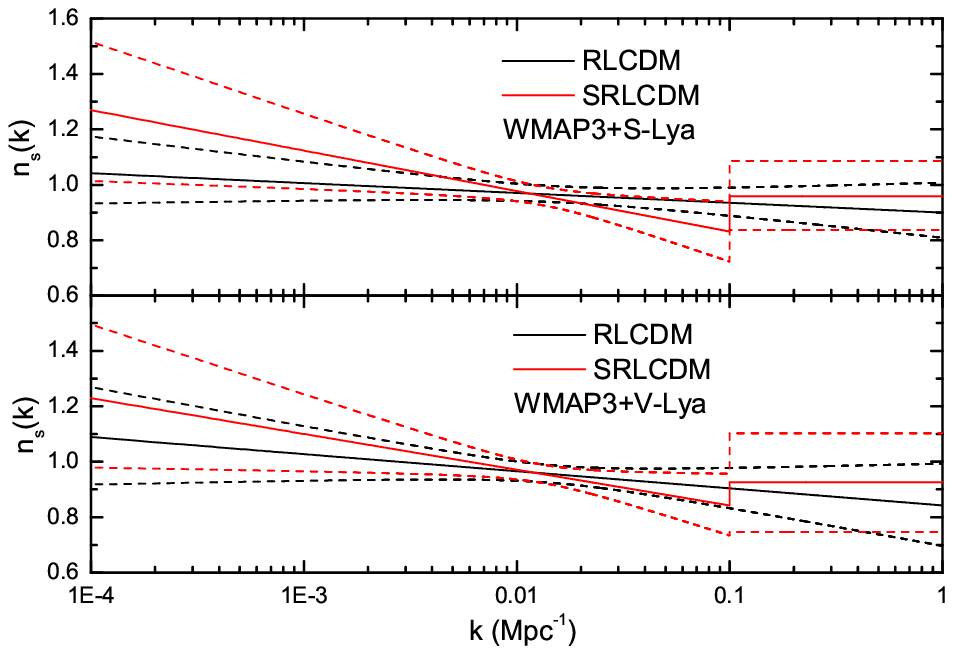}
\caption{Resulting 2$\sigma$ posterior constraints on $n_s(k)$ for
Running Spectral Index $\Lambda$CDM Model and Running Spectral Index
$\Lambda$CDM Model with step-like $n_s$ truncated at $k= 0.1$
Mpc$^{-1}$, as measured by WMAP3 in combination with LUQAS+CROFT
sample of Lyman $\alpha$ as analyzed in Ref. \cite{Viel:2004bf}
 and the SDSS Lyman $\alpha$ sample from
Ref. \cite{McDonald:2004xn}. The solid lines depict the median
intervals and the dashed lines are the 2$\sigma$ limits. HST, BBN
and age priors have been adopted. \label{fig:fig0012}}
\end{center}
\end{figure*}

In Fig.~\ref{fig:fig0012} we delineate the resulting 2$\sigma$
posterior constraints on $n_s(k)$ for Running Spectral Index
$\Lambda$CDM Model and Running Spectral Index $\Lambda$CDM Model
with step-like $n_s$ truncated at $k= 0.1$ Mpc$^{-1}$,  measured
by WMAP3 in combination with LUQAS+CROFT sample of Lyman $\alpha$ as
analyzed in Ref. \cite{Viel:2004bf}
 and the SDSS Lyman $\alpha$ sample from
Ref. \cite{McDonald:2004xn}.  One can easily find the scale
dependence on either of the upper or the bottom figures for the case
of SRLCDM model.

\section{discussion and conclusion}

 Through our detailed investigations we have found that values of
 cosmological parameters obtained by MCMC change considerably depending
 on whether we use low multipole components of CMB or not.
Currently, the WMAP team use a very large number of pixels
(666+1172 pixels for low multipole temperature and polarization
spectra \cite{Spergel:2006hy,Page:2006hz,Hinshaw:2006,WMAP3IE}),
which may take too large a weight for the determination of the
likelihood and hence cosmological parameters.
This may be somewhat problematic because these low multipoles suffer
from relatively large cosmic variance.
On the other hand, if we incorporate low $\ell$ polarization components,
the optical depth $\tau$ can be determined with a smaller ambiguity
because the degeneracy between $n_s$ and $\tau$ is partially removed,
which in turn result in a better determination of the spectral index.
Hence it is difficult to conclude which approach gives more precise
results.

We obtain different values of cosmological parameters with different
datasets.  For example, we find a very low $H_0$ in the combination
of WMAP3 (without $\ell \le 23$) and SDSS is rather interesting,
 even though we have adopted the HST
prior in Table 6.  Nonetheless these discrepancies are typically within
2$\sigma$ and do not imply the datasets we have used are incompatible
with each other, including the differences between
the Riess ``gold'' sample and the first year SNLS,
SDSS and 2dF, or
LUQAS+CROFT sample of Lyman $\alpha$ as
analyzed in Ref. \cite{Viel:2004bf} and the SDSS Lyman $\alpha$
sample from Ref. \cite{McDonald:2004xn}

While our results as a whole support the concordance cosmology, they
also indicate we are still far from being able to determine all of the
six cosmological parameters with two digits' accuracy in the
context of this simple power-law $\Lambda$CDM model.
Hence we should be open-minded to new physics which may be hidden in the
error bars of the current observational data and continue blind analysis
without any theoretical prejudices.  A good example is the possible
running of the spectral index, which is preferred to more than 2$\sigma$
in a number of combinations of datasets.

{\bf{Acknowledgments:}} We acknowledge the use of the Legacy Archive
for Microwave Background Data Analysis (LAMBDA). Support for LAMBDA
is provided by the NASA Office of Space Science. We have performed
our numerical analysis on the Shanghai Supercomputer Center (SSC).
We used a modified version of CAMB \cite{Lewis:1999bs,IEcamb} which
is based on CMBFAST \cite{cmbfast,IEcmbfast}. We thank Patrick
Mcdonald and Anze Slosar for kind discussions regarding their online
codes for SDSS Lyman alpha fittings \cite{IESlya1,IESlya2}. We are
grateful to Kohji Yoshikawa and Ting Zhang for computational
assistance. We thank Kevork Abazajian, Long-long Feng, Julien
Lesgourgues, Antony Lewis, Hiranya Peiris, David Spergel, Yasushi
Suto, Masahiro Takada, Max Tegmark, Matteo Viel, Yongzhong Xu,
Pengjie Zhang, Xinmin Zhang and Gongbo-Zhao for helpful discussions.
The work of J.Y. is supported partially by the JSPS Grant-in-Aid for
Scientific Research No. 16340076 and B.F. is by the JSPS fellowship
program. This work is supported in part by National Natural Science
Foundation of China under Grant Nos. 90303004, 10533010 and 19925523
and by Ministry of Science and Technology of China under Grant No.
NKBRSF G19990754.

{\it Note Added: } During the finalizing of our manuscript, Refs.
\cite{Eriksen:2006xr,Huffenberger:2006xx,Bridges:2006zm,Finelli:2006fi}
appeared on the arXivs which are somewhat relevant to our study.


\begin{thebibliography}{}

\bibitem{Sato:1980yn}
  K.~Sato,
  ``First Order Phase Transition Of A Vacuum And Expansion Of The Universe,''
  Mon.\ Not.\ Roy.\ Astron.\ Soc.\  {\bf 195}, 467 (1981).

\bibitem{Guth:1980zm}
  A.~H.~Guth,
  ``The Inflationary Universe: A Possible Solution To The Horizon And Flatness
  Problems,''
  Phys.\ Rev.\ D {\bf 23}, 347 (1981).

\bibitem{Linde:2005ht}
For a review of inflation, see e.g. A.~D.~Linde,
  ``Particle Physics and Inflationary Cosmology,''
  Contemp.\ Concepts Phys.\  {\bf 5}, 1 (2005)
  [arXiv:hep-th/0503203].

\bibitem{Bennett:2003bz}
  C.~L.~Bennett {\it et al.},
  ``First Year Wilkinson Microwave Anisotropy Probe (WMAP) Observations:
  Preliminary Maps and Basic Results,''
  Astrophys.\ J.\ Suppl.\  {\bf 148}, 1 (2003)
  [arXiv:astro-ph/0302207].

\bibitem{Kogut:2003et}
  A.~Kogut {\it et al.},
  ``Wilkinson Microwave Anisotropy Probe (WMAP) First Year Observations: TE
  Polarization,''
  Astrophys.\ J.\ Suppl.\  {\bf 148}, 161 (2003)
  [arXiv:astro-ph/0302213].

\bibitem{Peiris:2003ff}
  H.~V.~Peiris {\it et al.},
  ``First year Wilkinson Microwave Anisotropy Probe (WMAP) observations:
  Implications for inflation,''
  Astrophys.\ J.\ Suppl.\  {\bf 148}, 213 (2003)
  [arXiv:astro-ph/0302225].

\bibitem{Spergel:2003cb}
  D.~N.~Spergel {\it et al.}  [WMAP Collaboration],
  ``First Year Wilkinson Microwave Anisotropy Probe (WMAP) Observations:
  Determination of Cosmological Parameters,''
  Astrophys.\ J.\ Suppl.\  {\bf 148}, 175 (2003)
  [arXiv:astro-ph/0302209].

\bibitem{Hinshaw:2003ex}
  G.~Hinshaw {\it et al.},
  ``First Year Wilkinson Microwave Anisotropy Probe (WMAP) Observations:
  Angular Power Spectrum,''
  Astrophys.\ J.\ Suppl.\  {\bf 148}, 135 (2003)
  [arXiv:astro-ph/0302217].

\bibitem{Verde:2003ey}
  L.~Verde {\it et al.},
  ``First Year Wilkinson Microwave Anisotropy Probe (WMAP) Observations:
  Parameter Estimation Methodology,''
  Astrophys.\ J.\ Suppl.\  {\bf 148}, 195 (2003)
  [arXiv:astro-ph/0302218].



\bibitem{Spergel:2006hy}
  D.~N.~Spergel {\it et al.},
  ``Wilkinson Microwave Anisotropy Probe (WMAP) Three Year Results:
  Implications for Cosmology,''
  arXiv:astro-ph/0603449.



\bibitem{Page:2006hz} L.~Page {\it et al.},
  ``Three Year Wilkinson Microwave Anisotropy Probe (WMAP) Observations:
  Polarization Analysis,''
  arXiv:astro-ph/0603450.

\bibitem{Hinshaw:2006}
 G.~Hinshaw {\it et al.},
  ``Three-Year Wilkinson Microwave Anisotropy Probe (WMAP) Observations:
  Temperature Analysis,''
  arXiv:astro-ph/0603451.
\bibitem{Jarosik:2006}
 N.~Jarosik {\it et al.},
  ``Three-Year Wilkinson Microwave Anisotropy Probe (WMAP) Observations: Beam
  Profiles, Data Processing, Radiometer Characterization and Systematic Error
  Limits,''
  arXiv:astro-ph/0603452.

\bibitem{WMAP3IE}
Available at http://lambda.gsfc.nasa.gov/product/map/ .





\bibitem{Montroy:2005yx}
  T.~E.~Montroy {\it et al.},
  ``A Measurement of the CMB  Spectrum from the 2003 Flight of BOOMERANG,''
  arXiv:astro-ph/0507514.

\bibitem{Kuo:2002ua}
  C.~L.~Kuo {\it et al.}  [ACBAR collaboration],
  ``High Resolution Observations of the CMB Power Spectrum with ACBAR,''
  Astrophys.\ J.\  {\bf 600}, 32 (2004)
  [arXiv:astro-ph/0212289].


\bibitem{Readhead:2004gy}
  A.~C.~S.~Readhead {\it et al.},
  ``Extended Mosaic Observations with the Cosmic Background Imager,''
  Astrophys.\ J.\  {\bf 609}, 498 (2004)
  [arXiv:astro-ph/0402359].

\bibitem{Dickinson:2004yr} 
  C.~Dickinson {\it et al.},
  ``High sensitivity measurements of the CMB power spectrum with the extended
  Very Small Array,''
  Mon.\ Not.\ Roy.\ Astron.\ Soc.\  {\bf 353}, 732 (2004)
  [arXiv:astro-ph/0402498].


\bibitem{Cole:2005sx}
  S.~Cole {\it et al.}  [The 2dFGRS Collaboration],
  ``The 2dF Galaxy Redshift Survey: Power-spectrum analysis of the final
  dataset and cosmological implications,''
  Mon.\ Not.\ Roy.\ Astron.\ Soc.\  {\bf 362} (2005) 505
  [arXiv:astro-ph/0501174].

\bibitem{Tegmark:2003uf}
  M.~Tegmark {\it et al.}  [SDSS Collaboration],
  ``The 3D power spectrum of galaxies from the SDSS,''
  Astrophys.\ J.\  {\bf 606}, 702 (2004)
  [arXiv:astro-ph/0310725].


\bibitem{Riess:2004nr}
  A.~G.~Riess {\it et al.}  [Supernova Search Team Collaboration],
  ``Type Ia Supernova Discoveries at $z>1$ From the Hubble Space Telescope:
  Evidence for Past Deceleration and Constraints on Dark Energy Evolution,''
  Astrophys.\ J.\  {\bf 607}, 665 (2004)
  [arXiv:astro-ph/0402512].


\bibitem{snls}
  P.~Astier {\it et al.},
  ``The Supernova Legacy Survey: Measurement of $\Omega_M$, $\Omega_{\Lambda}$ and w
  from the First Year Data Set,''
  Astron.\ Astrophys.\  {\bf 447}, 31 (2006)
  [arXiv:astro-ph/0510447].


\bibitem{Wang:2002hf}
  X.~Wang, B.~Feng, M.~Li, X.~L.~Chen and X.~Zhang,
  ``Natural inflation, Planck scale physics and oscillating primordial
  spectrum,''
  Int.\ J.\ Mod.\ Phys.\ D {\bf 14}, 1347 (2005)
  [arXiv:astro-ph/0209242].


\bibitem{Feng:2003nt}
  B.~Feng, X.~Gong and X.~Wang,
  ``Assessing the Effects of the Uncertainty in Reheating Energy Scale on
  Primordial Spectrum and CMB,''
  Mod.\ Phys.\ Lett.\ A {\bf 19}, 2377 (2004)
  [arXiv:astro-ph/0301111].


\bibitem{Lewis:2002ah}
  A.~Lewis and S.~Bridle,
  ``Cosmological parameters from CMB and other data: a Monte-Carlo approach,''
  Phys.\ Rev.\ D {\bf 66}, 103511 (2002)
  [arXiv:astro-ph/0205436].

\bibitem{Cline:2003ve}
  J.~M.~Cline, P.~Crotty and J.~Lesgourgues,
  ``Does the small CMB quadrupole moment suggest new physics?,''
  JCAP {\bf 0309}, 010 (2003)
  [arXiv:astro-ph/0304558].


\bibitem{Efstathiou:2003tv}
  See  e.g.  G.~Efstathiou,
  Mon.\ Not.\ Roy.\ Astron.\ Soc.\  {\bf 348}, 885 (2004)
  [arXiv:astro-ph/0310207].


\bibitem{Slosar:2004fr}
  See  also A.~Slosar, U.~Seljak and A.~Makarov,
  ``Exact likelihood evaluations and foreground marginalization in low
  resolution WMAP data,''
  Phys.\ Rev.\ D {\bf 69}, 123003 (2004)
  [arXiv:astro-ph/0403073].


\bibitem{Feng:2003mk}
  B.~Feng, M.~z.~Li, R.~J.~Zhang and X.~m.~Zhang,
  ``An inflation model with large variations in spectral index,''
  Phys.\ Rev.\ D {\bf 68}, 103511 (2003)
  [arXiv:astro-ph/0302479].


\bibitem{Kawasaki:2003zv}
M.~Kawasaki, M.~Yamaguchi, and J.~Yokoyama, ``Inflation with a
running spectral index in supergravity,'' Phys.\ Rev.\ D {\bf 68},
023508 (2003) [arXiv:hep-ph/0304161].

\bibitem{Huang:2003zp}
  Q.~G.~Huang and M.~Li,
  ``CMB power spectrum from noncommutative spacetime,''
  JHEP {\bf 0306}, 014 (2003)
  [arXiv:hep-th/0304203].


\bibitem{Yamaguchi:2003fp}
M.~Yamaguchi and J.~Yokoyama, ``Chaotic hybrid new inflation in
supergravity with a running spectral index,'' Phys.\ Rev.\ D {\bf
68}, 123520 (2003) [arXiv:hep-ph/0307373].

\bibitem{Yamaguchi:2004tn}
M.~Yamaguchi and J.~Yokoyama, ``Smooth hybrid inflation in
supergravity with a running spectral index  and early star
formation,'' Phys.\ Rev.\ D {\bf 70}, 023513 (2004)
[arXiv:hep-ph/0402282].


\bibitem{Chen:2004nx}
C.-Y.~Chen, B.~Feng, X.-L.~Wang, and Z.-Y.~Yang, ``Reconstructing
large running-index inflaton potentials'', Class.\ Quant.\ Grav.\
{\bf 21}, 3223 (2004) arXiv:[astro-ph/0404419].

\bibitem{Yokoyama:1998rw}
  J.~Yokoyama,
  ``Chaotic new inflation and primordial spectrum of adiabatic fluctuations,''
  Phys.\ Rev.\ D {\bf 59}, 107303 (1999).

\bibitem{Yokoyama:1998pt}
  J.~Yokoyama,
  ``Chaotic new inflation and formation of primordial black holes,''
  Phys.\ Rev.\ D {\bf 58}, 083510 (1998)
  [arXiv:astro-ph/9802357].





\bibitem{Easther:2006tv}
  R.~Easther and H.~Peiris,
 ``Implications of a running spectral index for slow roll inflation,''
  arXiv:astro-ph/0604214.

\bibitem{Feng:2006zj}
  B.~Feng, J.~Q.~Xia, J.~Yokoyama, X.~Zhang and G.~B.~Zhao,
  ``Weighing neutrinos in the presence of a running primordial spectral
  index,''
  arXiv:astro-ph/0605742.
\bibitem{toappear}
B. Feng {\it et al.}, to appear.




\bibitem{Viel:2004bf}
  T.~S.~Kim, M.~Viel, M.~G.~Haehnelt, R.~F.~Carswell and S.~Cristiani,
  ``The power spectrum of the flux distribution in the Lyman-alpha forest of a
  Large sample of UVES QSO Absorption Spectra (LUQAS),''
  Mon.\ Not.\ Roy.\ Astron.\ Soc.\  {\bf 347}, 355 (2004)
  [arXiv:astro-ph/0308103];
  R.~A.~C.~Croft {\it et al.},
  ``Towards a Precise Measurement of Matter Clustering: Lyman-alpha Forest Data
  at Redshifts 2-4,''
  Astrophys.\ J.\  {\bf 581}, 20 (2002)
  [arXiv:astro-ph/0012324];
  M.~Viel, M.~G.~Haehnelt and V.~Springel,
  ``Inferring the dark matter power spectrum from the Lyman-alpha forest in
  high-resolution QSO absorption spectra,''
  Mon.\ Not.\ Roy.\ Astron.\ Soc.\  {\bf 354}, 684 (2004)
  [arXiv:astro-ph/0404600].



\bibitem{McDonald:2004xn}
P.~McDonald {\it et al.},
  ``The Lyman-alpha Forest Power Spectrum from the Sloan Digital Sky Survey,''
  Astrophys.\ J.\ Suppl.\  {\bf 163}, 80 (2006)
  [arXiv:astro-ph/0405013];
  P.~McDonald {\it et al.},
  ``The Linear Theory Power Spectrum from the Lyman-alpha Forest in the Sloan
  Digital Sky Survey,''
  Astrophys.\ J.\  {\bf 635}, 761 (2005)
  [arXiv:astro-ph/0407377].

\bibitem{IEMCMC}
Available at \texttt{http://cosmologist.info} .

\bibitem{Freedman:2000cf}
  W.~L.~Freedman {\it et al.},
  ``Final Results from the Hubble Space Telescope Key Project to Measure the
  Hubble Constant,''
  Astrophys.\ J.\  {\bf 553}, 47 (2001)
  [arXiv:astro-ph/0012376].


\bibitem{Burles:2000zk}
  S.~Burles, K.~M.~Nollett and M.~S.~Turner,
  ``Big-Bang Nucleosynthesis Predictions for Precision Cosmology,''
  Astrophys.\ J.\  {\bf 552}, L1 (2001)
  [arXiv:astro-ph/0010171].

\bibitem{sdssfit}
  M.~Tegmark {\it et al.}  [SDSS Collaboration],
  ``Cosmological parameters from SDSS and WMAP,''
  Phys.\ Rev.\ D {\bf 69}, 103501 (2004)
  [arXiv:astro-ph/0310723].



\bibitem{DiPietro:2002cz}
For details see e.g.
  E.~Di Pietro and J.~F.~Claeskens,
  ``Quintessence models faced with future supernovae data,''
  Mon.\ Not.\ Roy.\ Astron.\ Soc.\  {\bf 341}, 1299 (2003),
  [arXiv:astro-ph/0207332].

\bibitem{Prieto:2006ex}
  J.~L.~Prieto, A.~Rest and N.~B.~Suntzeff,
  ``A New Method to Calibrate the Magnitudes of Type Ia Supernovae at Maximum
  Light,''
  arXiv:astro-ph/0603407.

\bibitem{Jha:2002af}
  S.~Jha,
  ``Exploding stars, near and far,'' PhD thesis, Harvard University.

\bibitem{Guy:2005me}
  J.~Guy, P.~Astier, S.~Nobili, N.~Regnault and R.~Pain,
  ``SALT: a Spectral Adaptive Light curve Template for Type Ia Supernovae,''
  arXiv:astro-ph/0506583.

\bibitem{Wang:2005qn}
  X.~F.~Wang, L.~F.~Wang, X.~Zhou, Y.~Q.~Lou and Z.~W.~Li,
  ``A Novel Color Parameter As A Luminosity Calibrator for Type Ia
  Supernovae,''
  Astrophys.\ J.\  {\bf 620}, L87 (2005)
  [arXiv:astro-ph/0501565].

\bibitem{Wang:2006cq}
  X.~F.~Wang, L.~F.~Wang, R.~Pain, X.~Zhou and Z.~W.~Li,
  ``Determination of the Hubble constant, the intrinsic scatter of luminosities
  of Type Ia SNe, and evidence for non-standard dust in other galaxies,''
  arXiv:astro-ph/0603392.


\bibitem{Wang:06ap}
Available at http://sn2006.pmo.ac.cn/; X.~F.~Wang {\it et al.}, to
appear.



\bibitem{Viel:2006yh}
  M.~Viel, M.~G.~Haehnelt and A.~Lewis,
  ``The Lyman-alpha forest and WMAP year three,''
  arXiv:astro-ph/0604310.


\bibitem{Seljak:2006bg}
  U.~Seljak, A.~Slosar and P.~McDonald,
  ``Cosmological parameters from combining the Lyman-alpha forest with CMB,
  galaxy clustering and SN constraints,''
  arXiv:astro-ph/0604335.

\bibitem{Lesgourgues:2006nd}
 See  e.g.  J.~Lesgourgues and S.~Pastor,
  ``Massive neutrinos and cosmology,''
  arXiv:astro-ph/0603494.

\bibitem{IESlya1}
Available at  http://www.cita.utoronto.ca/pmcdonal/LyaF/ .

\bibitem{IESlya2}
Available at http://www.slosar.com/aslosar/lya.html  .


\bibitem{Contaldi:2003zv}
  C.~R.~Contaldi, M.~Peloso, L.~Kofman and A.~Linde,
  ``Suppressing the lower Multipoles in the CMB Anisotropies,''
  JCAP {\bf 0307}, 002 (2003)
  [arXiv:astro-ph/0303636].


\bibitem{Dore:2003wp}
  O.~Dore, G.~P.~Holder and A.~Loeb,
  ``The CMB Quadrupole in a Polarized Light,''
  Astrophys.\ J.\  {\bf 612} (2004) 81
   [arXiv:astro-ph/0309281].


\bibitem{Bridle:2003sa}
  S.~L.~Bridle, A.~M.~Lewis, J.~Weller and G.~Efstathiou,
  ``Reconstructing the primordial power spectrum,''
  Mon.\ Not.\ Roy.\ Astron.\ Soc.\  {\bf 342}, L72 (2003)
  [arXiv:astro-ph/0302306].

\bibitem{Bond:1987ub}
  J.~R.~Bond and G.~Efstathiou,
  ``The Statistics Of Cosmic Background Radiation Fluctuations,''
  Mon.\ Not.\ Roy.\ Astron.\ Soc.\  {\bf 226}, 655 (1987).



\bibitem{Hodges:1989dw}
  H.~M.~Hodges, G.~R.~Blumenthal, L.~A.~Kofman and J.~R.~Primack,
  ``Nonstandard Primordial Fluctuations From A Polynomial Inflaton Potential,''
  Nucl.\ Phys.\ B {\bf 335}, 197 (1990).

\bibitem{Matsumiya:2001xj}
  M.~Matsumiya, M.~Sasaki and J.~Yokoyama,
  ``Cosmic inversion: Reconstructing primordial spectrum from CMB
  anisotropy,''
  Phys.\ Rev.\ D {\bf 65}, 083007 (2002)
  [arXiv:astro-ph/0111549].

\bibitem{Matsumiya:2002tx}
  M.~Matsumiya, M.~Sasaki and J.~Yokoyama,
  ``Cosmic Inversion II --An iterative method for reproducing the primordial
  spectrum from the CMB data--,''
  JCAP {\bf 0302}, 003 (2003)
  [arXiv:astro-ph/0210365].

\bibitem{Kogo:2003yb}
  N.~Kogo, M.~Matsumiya, M.~Sasaki and J.~Yokoyama,
  ``Reconstructing the primordial spectrum from WMAP data by the cosmic
  inversion method,''
  Astrophys.\ J.\  {\bf 607}, 32 (2004)
  [arXiv:astro-ph/0309662].

\bibitem{Kogo:2004vt}
  N.~Kogo, M.~Sasaki and J.~Yokoyama,
  ``Reconstructing the Primordial Spectrum with CMB Temperature and
  Polarization,''
  Phys.\ Rev.\ D {\bf 70}, 103001 (2004)
  [arXiv:astro-ph/0409052].

\bibitem{Kogo:2005qi}
  N.~Kogo, M.~Sasaki and J.~Yokoyama,
  ``Constraining cosmological parameters by the cosmic inversion method,''
  Prog.\ Theor.\ Phys.\  {\bf 114}, 555 (2005)
  [arXiv:astro-ph/0504471].


\bibitem{Lewis:2006ma}
  A.~Lewis,
  ``Observational constraints and cosmological parameters,''
  arXiv:astro-ph/0603753.

\bibitem{GR92} A. Gelman and
D. Rubin, `` Inference from iterative simulation using multiple
sequences,'' Statistical Science {\bf 7}, 457 (1992).


\bibitem{Lewis:1999bs}
   A.~Lewis, A.~Challinor and A.~Lasenby,
  ``Efficient Computation of CMB anisotropies in closed FRW models,''
  Astrophys.\ J.\  {\bf 538}, 473 (2000)
  [arXiv:astro-ph/9911177].


\bibitem{IEcamb}
Available at http://camb.info  .

\bibitem{cmbfast}
  U.~Seljak and M.~Zaldarriaga,
  ``A Line of Sight Approach to Cosmic Microwave Background Anisotropies,''
  Astrophys.\ J.\  {\bf 469}, 437 (1996)
  [arXiv:astro-ph/9603033].

\bibitem{IEcmbfast}
Available at http://cmbfast.org/   .


\bibitem{Eriksen:2006xr}
  H.~K.~Eriksen {\it et al.},
   ``A re-analysis of the three-year WMAP temperature power spectrum and
  arXiv:astro-ph/0606088.

\bibitem{Huffenberger:2006xx}
  K.~M.~Huffenberger, H.~K.~Eriksen and F.~K.~Hansen,
   ``Point source power in three-year Wilkinson Microwave Anisotropy Probe
  data,''
  arXiv:astro-ph/0606538.

\bibitem{Bridges:2006zm}
  M.~Bridges, A.~N.~Lasenby and M.~P.~Hobson,
  ``WMAP 3-year primordial power spectrum,''
  arXiv:astro-ph/0607404.

\bibitem{Finelli:2006fi}
  F.~Finelli, M.~Rianna and N.~Mandolesi,
  ``WMAP Three Year Constraints on the Inflationary Expansion,''
  arXiv:astro-ph/0608277.



\end{thebibliography}
\end{document}